\pdfoutput=1
\documentclass[paper=a4, fontsize=12pt]{scrartcl}
\usepackage[utf8x]{inputenc}				
\usepackage{lmodern} 
\usepackage[english]{babel}
\usepackage[intlimits]{amsmath}
\usepackage{amssymb,amstext,dsfont}
\usepackage[pdftex]{graphicx}
\usepackage{color}
\usepackage{xfrac} 
\usepackage{braket}
\usepackage{cite}

\newif\ifdraft

\DeclareGraphicsExtensions{.pdf, .png, .bmp, .jpg}

\newcommand{\norm}{\mathcal{N}}

\newcommand{\upd}{\mathrm{d}}

\newcommand{\ints}[3]{\int_{#1}^{#2}\!\upd #3}
\newcommand{\iu}{\mathrm{i}}
\newcommand{\eu}{\mathrm{e}}

\newcommand{\Tr}{\operatorname{Tr}}
\newcommand{\tT}{T} 
\newcommand{\eg}{{\em e.g. }}
\newcommand{\ie}{{\em i.e. }}
\newcommand{\cf}{{\em cf. }}
\newcommand{\teq}{\! = \!}

\newcommand{\eps}{\epsilon}
\newcommand{\lamb}{\lambda}
\newcommand{\xe}{\eps_\text{min}}
\newcommand{\ye}{\eps_\text{max}}
\newcommand{\xl}{\lamb_\text{min}}
\newcommand{\yl}{\lamb_\text{max}}

\newcommand{\emin}{\xe}
\newcommand{\emax}{\ye}
\newcommand{\lmin}{\xl}
\newcommand{\lmax}{\yl}

\newcommand{\Foo}{F_{kl}^{(0)}}
\newcommand{\Fot}{F_{kl}^{(\lamb)}}
\newcommand{\Fto}{F_{kl}^{(\eps)}}
\newcommand{\Ftt}{F_{kl}^{(\eps\lamb)}}
\newcommand{\nl}{n_{\lt}}
\newcommand{\nk}{n_{\kt}}
\newcommand{\kt}{\tilde{k}}
\newcommand{\lt}{\tilde{l}}
\newcommand{\spert}{\frac{N}{2\nu}}
\newcommand{\dte}{\delta_{\eps}}
\newcommand{\dtl}{\delta_{\lamb}}
\newcommand{\pin}{p_\nu^\text{in}}
\newcommand{\pex}{p_\nu^\text{ex}}
\newcommand{\pgue}{p^\text{GUE}}
\newcommand{\pw}{p^\text{W}}
\newcommand{\ps}{p^\text{s}}
\newcommand{\pnu}{p_\nu}
\newcommand{\cin}{c_{\text{in}}}
\newcommand{\cex}{c_{\text{ex}}}

\newcommand{\plr}{p^{(\eps\lamb)}}

\title{Spectral statistics of nearly unidirectional quantum graphs}
\author{Maram Akila and Boris Gutkin\footnote{maram.akila@uni-due.de,  boris.gutkin@uni-due.de -- 
Faculty of Physics, University of Duisburg-Essen, Lotharstr. 1, 47048 Duisburg, Germany }}

\draftfalse
\begin{document}
\maketitle
\begin{abstract}
\textbf{Abstract:} The energy levels of a quantum graph with time reversal symmetry and  unidirectional classical dynamics are doubly degenerate and obey the spectral statistics of the Gaussian Unitary Ensemble.  These degeneracies, however,  are lifted when  the unidirectionality is broken in one of the graph's vertices by a singular perturbation. Based on a Random Matrix model   we derive an analytic expression for the nearest neighbour distribution between energy levels of such systems.
As we demonstrate the result agrees excellently with the actual statistics for graphs with a uniform distribution of eigenfunctions. Yet, it exhibits quite substantial deviations for classes of graphs which show strong scarring.
\end{abstract}
Pacs: {02.10.Ox, 02.70.Hm, 03.65.Ge, 03.65.Sq, 05.45.Mt}\\
Keywords: {Quantum graphs, RMT, Anomalous spectral statistics}


\section{Introduction}
\label{sec:intro}

Chaotic quantum systems generically fall into three different universality  classes whose local spectral statistics are equivalent to those of Gaussian Ensembles \cite{dyson} of random Hermitian matrices. They are prescribed by general symmetry properties of the underlying Hamiltonians. In the case of a spinless, time reversal invariant (\textit{TRI}) system one normally expects, on the scales of mean level spacing (\textit{MLS}), spectral  properties corresponding to a Gaussian Orthogonal Ensemble (\textit{GOE}). If TRI is broken instead a Gaussian Unitary Ensemble (\textit{GUE}) describes the local spectral statistics. Finally, if the system has spin degrees of freedom, and TRI is not broken, its spectral statistics  belong to the symplectic class (\textit{GSE}).

\begin{figure}[bhtp]
\centering
\includegraphics[width=0.9\textwidth]{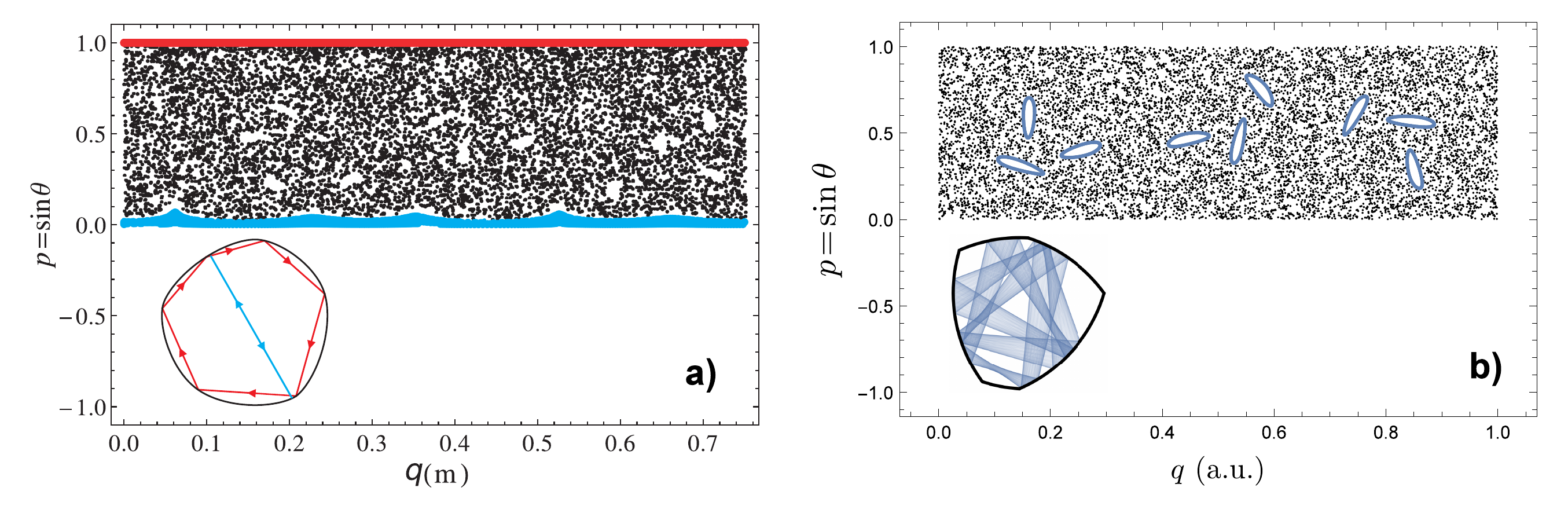}
\caption{\textit{(Colour online)} Shown are the classical phase spaces of two constant width billiards with in case (a) smooth boundaries (figure taken from \cite{bDietz}) and in (b) with corners.
Due to the unidirectionality of the systems trajectories in the upper part of the phase-space can not access the lower half, yet as the systems posses TRI the not shown halves look identical. On the left hand side the smooth boundaries give rise to bouncing ball modes (cyan) with adjacent KAM tori (cyan, \(p\!\approx\! 0\)) which separate the phase space into two components, for further information see \cite{bDietz}.
In contrast, the separating trajectories in the Reuleaux billiard, figure (b), are singular lines hitting the corners.}
\label{pic:billiardPS}
\end{figure}

It is well known that the  above universality classification might be affected by the presence of   additional symmetries  in the system \cite{berry, AnomalousSpectrum, sebastian}. 
However,  the universality class  might also change due to special dynamical properties. 
In particular, this happens in Hamiltonians with classically unidirectional chaotic dynamics, where two directions of motion are ergodically separated from one another \cite{boris1, boris2, prosen}. Such systems can be realized \eg in quantum billiards of constant width as  depicted in fig.~\ref{pic:billiardPS}. The first figure (a) shows an experimental realisation of a billiard with smooth boundaries from \cite{bDietz}. In this case the underlying spectrum was found to consist (primarily) of quasi-degenerate pairs whereas they don't obey the expected GOE statistics for TRI systems, but instead exhibit GUE behaviour. Qualitatively, the degeneracy can be understood from the fact that ``left-moving'' and  ``right-moving'' modes are dynamically separated. Although classically these two motions are ergodically disconnected, quantum mechanically the two modes are still weakly coupled due to dynamical tunnelling through an integrable region of KAM-tori around the bouncing ball modes.

\begin{figure}[hbtp]
\centering
\includegraphics[width=0.6\textwidth]{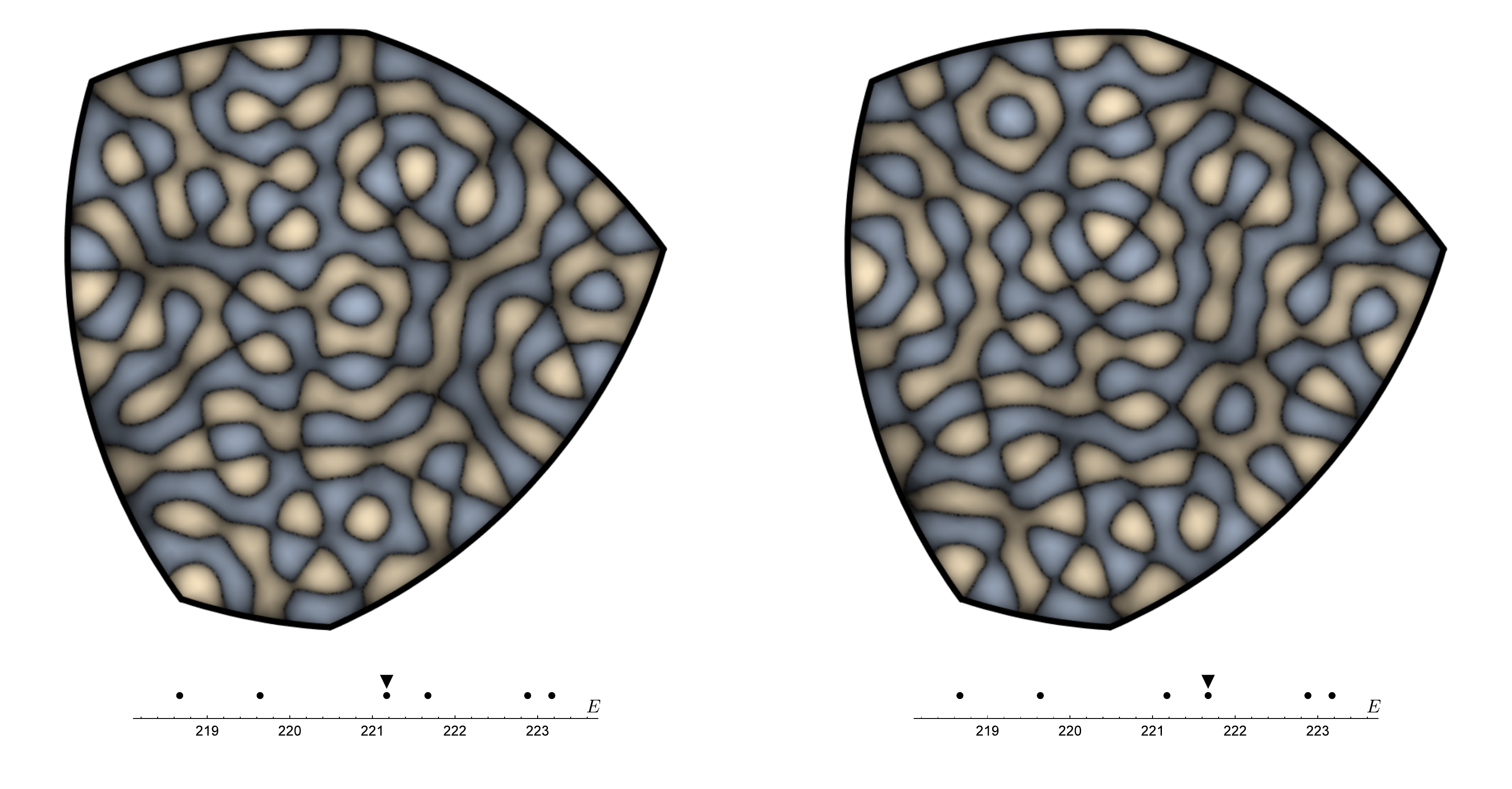}
\caption{\textit{(Colour online)} A doublet (eigenmodes 221 and 222) of the Reuleaux polygon whose classical phase space is shown in fig.~\ref{pic:billiardPS}(b). Dark areas correspond to negative values of the wavefunction, light to positive. Black ``lines'' in-between denote the nodal lines.
The energy scale below is given in units of MLS.}
\label{pic:relFace}
\end{figure}

For the second class, billiards with non-smooth boundaries such as the Reuleaux polygon shown in fig.~\ref{pic:billiardPS}(b), the splittings between quasi-degenerate states behave quite differently. In this case the dynamical barrier in the middle of the phase space shrinks to zero and the tunnelling occurs due to  diffractional orbits hitting the corners of the billiard domain.  In contrast to billiards with smooth boundaries this tunnelling effect is much stronger leading to  large  splittings  comparable to the mean level spacing.
See for instance fig.~\ref{pic:relFace} showing a doublet of the Reuleaux billiard.
The effect induced by the billiard  corners is reminiscent of singular perturbations;
 on the semiclassical level both cases give rise to quite similar, singular (diffractional) classical orbits.
Although it is known, see for instance \cite{mSieber,bogomolnyScatterer}, that normally neither the presence of corners nor singular perturbations   affects the spectral statistics, one can anticipate essentially different results for unidirectional billiards. Contributions arising from diffractional orbits  break down the  unidirectionality of the billiard  dynamics, thus effectively changing the universality class of the system. As a result, the spectral statistics of these systems do not belong to any of the three standard classes and it is the aim of this paper to study this phenomenon.

Rather than considering billiards 
we  will focus on the nearest neighbour distribution \( \pnu(s) \)  of  eigenvalues  in   quantum graphs with nearly  unidirectional classical ``dynamics''. They allow us to study the isolated effect of a singular perturbation with tunable strength \( \nu \). Furthermore, the rank of the singular perturbation can easily be adjusted by prescribing the number of vertices where the unidirectionality is broken.  In the present work we primarily study the effect of rank 1 perturbations. 

\section{Unidirectional Quantum Graphs}
\label{sec:qg1}

Quantum graphs are a widely used toy-model of chaotic quantum systems. We  will first sketch  below their general  properties  (for a more detailed account see for instance the review \cite{graphsReview} or \cite{qugaboo}) and then introduce the family of unidirectional quantum graphs.
 
A (closed) quantum graph consists of a set of \( B \) finite length \(l_j\) edges connected at \(V \) vertices. 
The edges  can be thought of as ideal 1D waveguides on which a wave function \( \psi \) propagates. In our case the propagation is free, \ie \( H\psi_j =-\mathop{{}\bigtriangleup} \nolimits \psi_j = k^2\psi_j \), therefore \( \psi_j=a_j\eu^{+\iu k x} +b_j\eu^{-\iu k x}\) at each edge \( j\). The true complexity of such systems stems from the boundary conditions at the graph's vertices.  At any given vertex \( i \)  we are faced with a number of incoming waves, forming the vector \( \vec{\psi}^{(i)}_\text{in} \), and an equal number of outgoing waves on the same edges forming the vector \( \vec{\psi}^{(i)}_\text{out} \).  The boundary condition describing the vertex is a unitary matrix \( \sigma_i \) matching both vectors,
\begin{equation}
\vec{\psi}^{(i)}_\text{out}
=\sigma_i\vec{\psi}^{(i)}_\text{in}\,.
\label{eq:locScatConcept}
\end{equation}
It is often referred to as ``local scattering'' matrix, and its unitarity ensures the conservation of local probability. 

Using the local scattering matrices, an internal scattering matrix \( S \) for the total graph can be constructed. It maps the vector of directed, incoming wave-function amplitudes
 \(  \vec{\Psi}_\text{in}= (\vec{\psi}^{(1)}_\text{in},\vec{\psi}_\text{in}^{(2)},\dots, \vec{\psi}^{(V)}_\text{in})^{\tT} \)
 onto the outgoing ones
 \(  \vec{\Psi}_\text{out}= (\vec{\psi}^{(1)}_\text{out},\vec{\psi}_\text{out}^{(2)},\dots, \vec{\psi}^{(V)}_\text{out})^\tT \).
For the eigenstates of the system the outgoing wave functions acquire the phases  \( \eu^{+\iu k l_j} \) during propagation along the respective edges and turn into incoming wave functions again. We can represent this by a self consistent equation
\begin{equation}
S\cdot\underbrace{\text{diag}{(\eu^{\iu k l_1}, \ldots, \eu^{\iu k l_{2B}})}}_{\mathcal{L}(k)}\;\vec{\Psi}_\text{out}=\vec{\Psi}_\text{out}
\,.
\label{eq:graphScatEv}
\end{equation}
As we describe every edge by two directions, each  length \( l_j\)  appears twice in \( \mathcal{L}(k) \).
The eigenenergies of the graph are now those \( k_n^2 \) for which we can find a \( \vec{\Psi}_\text{out} \) fulfilling above equation. In other words, the matrix \( 1- S \mathcal{L}(k) \) needs to have a zero eigenvalue at  \(k\teq k_n\).  If the lengths \( l_j \) are non commensurate, \ie they are given by rationally independent  numbers, one in general expects either GOE or GUE  statistics (depending on the symmetry   of the local scatterers)  for the graphs spectrum \cite{uzy}.

\begin{figure}[bt]
\centering
\includegraphics[width=1\textwidth]{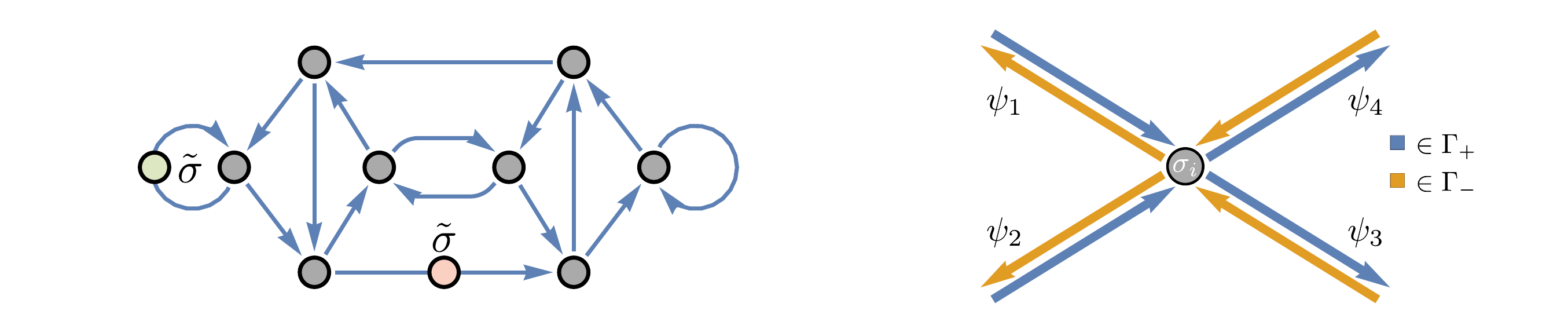}
\caption[Schematic drawing of unidirectional graphs]{\textit{(Colour online)} - 
\textit{Right hand side:} Illustration of a graph vertex with four attached leads which feature two separate directions, dark/blue and light/orange. Locally the scattering matrix does not couple those directions, \eg waves entering through an incoming orange edge only leave via an outgoing orange edge, as indicated in Equation (\ref{eq:locScatForm}). This ensures unidirectionality on a local scale.\\
\textit{Left hand side:}  Directed graph  \(\Gamma_+\) corresponding to a unidirectional De Bruijn quantum graph.  Its symmetric counterpart  \(\Gamma_-\) is obtained by reversing the direction of each edge. We highlighted two possible positions (light red/middle and light green/left) at which a backscattering vertex \(\tilde{\sigma}\), Equation (\ref{eq:backscatSigma}), could be placed. Note though that for a rank one perturbation only one  scatter can  be introduced. We comment on the differences of both positions in Section \ref{sec:comp}.}
\label{pic:scattererSchema}
\label{pic:unidirecGraphSchema}
\end{figure}

Starting from this general set-up only small adjustments are needed to introduce the family of unidirectional quantum graphs. The classical ``dynamics" on  a graph \(\Gamma\) can be thought of as free motion of a point-like particle on edges of the graph combined with   stochastic transitions  at  its vertices. The probability to pass from edge \(i\) to edge \(j\) is  defined by  the element \(|S_{i,j}|^2\) 
of the scattering matrix.    For a unidirectional graph we need to fix the vertex matrices  \(\sigma_i\) in  such a way that it would be impossible to switch the direction of motion along the edges.
To this end  we split an undirected graph  \(\Gamma\) into two directed ``halves"  \( \Gamma_\pm \) such that the n-th edges of \( \Gamma_+ \) and  \( \Gamma_- \) correspond to two possible directions of motions on the n-th edge of  \(\Gamma\). In addition, we require that the number of outgoing and incoming edges  at each vertex of  \( \Gamma_\pm \) would be identical. Note that such a splitting  of  \(\Gamma\) is  possible if and only if it possesses  an even number of edges per vertex (In such a case \(\Gamma\) admits euler cycles which can be used to assign the directions along the edges).
The following  structure of the vertex scattering matrices,
\begin{equation}
\sigma_i=\left(
\begin{array}{cc}
0 & U_i \\
U_i^\tT & 0 \\
\end{array}
\right)
\qquad
\text{with}\quad
U_i U_i^\dagger=U_i^\dagger U_i=1\,,
\label{eq:locScatForm}
\end{equation}
ensures that dynamics on  \( \Gamma_\pm\) are completely decoupled, see fig.~\ref{pic:scattererSchema}. Here we arranged \( \vec{\psi}^{(i)}_\text{in}\teq (\vec{\psi}^{(i+)}_\text{in},\vec{\psi}^{(i-)}_\text{in})^\tT \) such that  all entries \(\vec{\psi}^{(i+)}_\text{in}\) of \( \Gamma_+ \) are listed first, which makes the underlying block structure apparent.
Note  that all outgoing directions on \( \Gamma_+ \) are incoming directions of \( \Gamma_- \) and the other way around. Therefore, both directions are present in equal number making \( U_i \) a square matrix, which is a requirement to make \( U_i \) unitary.
Due to the off-diagonal structure  of  the \( \sigma \)'s the transition from graph  \( \Gamma_+ \) to  \( \Gamma_- \) (and vice versa) is impossible. In other words, a particle launched in one direction cannot switch to the opposite one.  
Further on choosing the same \( U_i \) for both blocks, implying \(\sigma_i \teq \sigma_i^\tT \), ensures that  the system possesses  TRI. The block structure in (\ref{eq:locScatForm}) already indicates that the corresponding quantum graph features an exactly double degenerate spectrum.
This becomes apparent when we cast \(S\) into a block-diagonal form. To this end we reorder the directed edges of \(\Gamma\) such that entries of \(\Gamma_+\) appear first, \ie \(\vec{\Psi}_\text{out}  \teq (\psi_+^{(1)},\dots,\psi_+^{(B)}, \psi_-^{(1)},\dots,\psi_-^{(B)})^\tT\). Within such order \(S\) takes the form:
\begin{equation}
 S= \left(
\begin{array}{cc}
\mathcal{S} & 0 \\
0 & \mathcal{S}^T \\
\end{array}
\right)\,.
\end{equation}
Both directions \(\Gamma_+\) and \(\Gamma_-\) form a chiral basis and are dynamically disconnected. Furthermore, as they are related via a transpose of \(\mathcal{S}\) we have \(l_j\teq l_{j+B}\) for the entries of the (also reordered) diagonal matrix \(\mathcal{L}(k)\).

\section{Breaking Unidirectionality}
\label{sec:rmt}
\label{sec:secEQ}

We will now consider graphs where unidirectionality is broken at just one  of the graph's vertices. To this end   we augment one of the edges of a unidirectional graph  \(\Gamma_0\)  with  a backscattering vertex. A general  \(2\times 2\) TRI backscattering matrix serving this purpose can be written as  
\begin{equation}
\tilde{\sigma}=\eu^{\iu \alpha+\iu\gamma} \left(
\begin{array}{cc}
r & t \\
t & -r^{*} \\
\end{array}
\right)
\qquad\text{with}\quad
r=\iu \eu^{\iu\beta}\sin{\alpha}\,,\quad
t=\cos{\alpha}\,,
\label{eq:backscatSigma}
\end{equation}
where \(*\) denotes complex conjugation. The parameter \( \alpha \) controls the strength of back-scattering:
In the case of \( \alpha\teq 0 \) the vertex becomes transparent with the back-reflection \( r\teq 0\) and all incoming waves pass through (acquiring a phase \(\eu^{\iu\gamma}\)).
In the opposite case, for \(\alpha\teq\sfrac{\pi}{2}\), the transmission \( t\) becomes zero and the degeneracies are lifted with the splittings reaching their maximum values. The parameter \(\beta\) allows back-scattered waves on \(\Gamma_\pm\) to acquire different phases.
But, the spectrum is unaffected by the choice of this parameter as can be inferred from semi-classical arguments. 
A trajectory starting on \(\Gamma_+\) (resp. \(\Gamma_-\)) on this nearly unidirectional graph needs to back-scatter twice at \(\tilde{\sigma}\) to return to \(\Gamma_+\) (\(\Gamma_-\)). This leads to a cancellation of the phase associated with \(\beta\) for all closed orbits.
The ``global'' phase \(\gamma\) leads to a shift of the spectrum which does not change the spectral statistics, \eg \(\pnu(s)\).

Given such an additional scatterer the  matrix \( S \) for the total graph can be written in the form
\begin{equation}
S=
S_0(\gamma)+(\eu^{2\iu\,\alpha}-1) S_0(\gamma) \ket{\psi}\bra{\psi}
 =S_0(\gamma)\cdot\exp \left(2\iu\,\alpha \ket{\psi}\bra{\psi}\right)
\label{eq:degS}
\end{equation}
\begin{equation}
\text{with}\qquad
\ket{\psi}=
\frac{\eu^{\iu\beta/2}}{\sqrt{2}} \ket{\text{in}_+}
+\frac{\eu^{-\iu\beta/2}}{\sqrt{2}} \ket{\text{in}_-}
\,, 
\label{eq:pertVec}
\end{equation}
Herein  \( S_0 \) is the unperturbed scattering matrix of a unidirectional  quantum graph, containing \( \tilde{\sigma} \) in its transparent form, \ie for \(\alpha\teq 0\), and $\ket{\psi}\bra{\psi}$ is a rank 1 perturbation. Please note that \( S_0 \) according to our definition depends on \(\gamma\).  Only for \(\gamma\teq 0\) the introduction of \(\tilde{\sigma}\) does not change the spectrum   of the original  graph \(\Gamma_0\).
The non-zero components of \(\ket{\text{in}_\pm}\) in eq.~(\ref{eq:pertVec}) correspond to the incoming directed edges on \(\Gamma_\pm \) leading to the scatterer. The action of \(S_0(\gamma)\) on these vectors should be understood in the form \( S_0(\gamma) \ket{\text{in}_\pm} \teq \eu^{\iu\gamma} \ket{\text{out}_\pm}\), \ie as mapping incoming wave functions onto outgoing ones. Note that instead of adding an additional vertex to the graph we  could  also change the local scattering matrix of a 4 edge vertex of   \(\Gamma_0\)  such that it  corresponds  to the standard Neumann boundary conditions (\(\sum_{i=1}^4\partial\psi_i\teq 0\)). In this case the decomposition (\ref{eq:degS}) 
 of \(S\) into ``unidirectional"  and ``singular" parts   holds as well. In fact, all the results of the present paper are applicable to quantum graphs possessing this particular form of scattering matrix. Due to the singular type of perturbation we call these graphs \textit{nearly unidirectional}.

\textbf{Secular Equation.} We can develop some systematic insight into the graph's spectral properties by looking at the eigenvalues of the unitary quantum evolution operator:
\begin{equation}
S \mathcal{L}(k) \ket{\lamb_m}= \eu^{\iu \lamb_m}\ket{\lamb_m}. 
\end{equation}
Using eq.~(\ref{eq:degS}) it is straightforward to obtain a secular equation which relates the eigenvalues \( \eu^{\iu \lamb_m}  \) of the perturbed system (\(\alpha\!\neq \! 0\)) to the  doubly degenerate eigenvalues \( \eu^{\iu \eps_m}  \) of the unperturbed one  (\(\alpha\teq 0\)). 
Expanding the eigenvectors  \( \ket{\lamb_m} \)   in terms of the old  \(\alpha\teq 0\) basis  \( \ket{\eps_m} \)  yields
\begin{equation}
\frac{-\iu\eu^{-\iu\alpha}}{2\sin{\alpha}}=
\sum_{m=1}^{B}\frac{\eu^{\iu \eps_m}|A_m|^2}{\eu^{\iu \lamb}-\eu^{\iu \eps_m}}\,,
\label{eq:secEqGraphComplex}
\end{equation}
where the left hand side depends only on \(\alpha\).
Solutions to this equation in \(\lamb\) provide the spectrum of eigenphases \(\{\lamb_m\}\) of the  perturbed system.
Importantly,  this equation only determines one half of the \(2B\) eigenvalues of the matrix \(S\mathcal{L}(k)\). The other half are pinned to their initial values \(\eps_m\)  for \(\alpha\teq 0\).  
Indeed, since the original spectrum of the unidirectional graph is doubly degenerate one half of all eigenvalues is not affected by a rank 1 perturbation. (A simple way to see this is to notice that the spectra of the original and the perturbed graphs satisfy the interlacing property, see below.)
The coefficients \( |A_m|^2 \teq |\braket{\eps_m | \text{in}_+}|^2 + |\braket{\eps_m | \text{in}_-}|^2 \) are the absolute values of the \( \ket{\eps_m} \) eigenvectors' component corresponding to the edge(s) where  \( \tilde{\sigma} \) is located.\footnote{For \(\alpha\teq 0\),   \( |A_m|^2 \) is identical on  both edges (\ie \(|\braket{\eps_m | \text{in}_\pm}|^2 \teq |\braket{\eps_m | \text{out}_\pm}|^2\)) and independent of the direction of \(\ket{\eps_m}\) (\ie whether \(\ket{\eps_m}\) resides on \(\Gamma_+\) or \(\Gamma_-\)). Yet, the incoming and outgoing amplitudes  are related via \( \exp{(\iu k l^\text{(in)})} A_m^\text{(in)} = \exp{(\iu \eps_m)} A_m^\text{(out)} \).  This explains the appearance of the phase factor in the numerator of eq.~(\ref{eq:secEqGraphComplex}).}
In the physically interesting cases where one of the eigenvalues \(\lambda\) is 0, \ie eq.~(\ref{eq:graphScatEv}) holds, the corresponding \( A_m \) are the amplitudes of the stationary wave solution on the graph.

\begin{figure}[hbtp]
\centering
\includegraphics[width=0.35\textwidth]{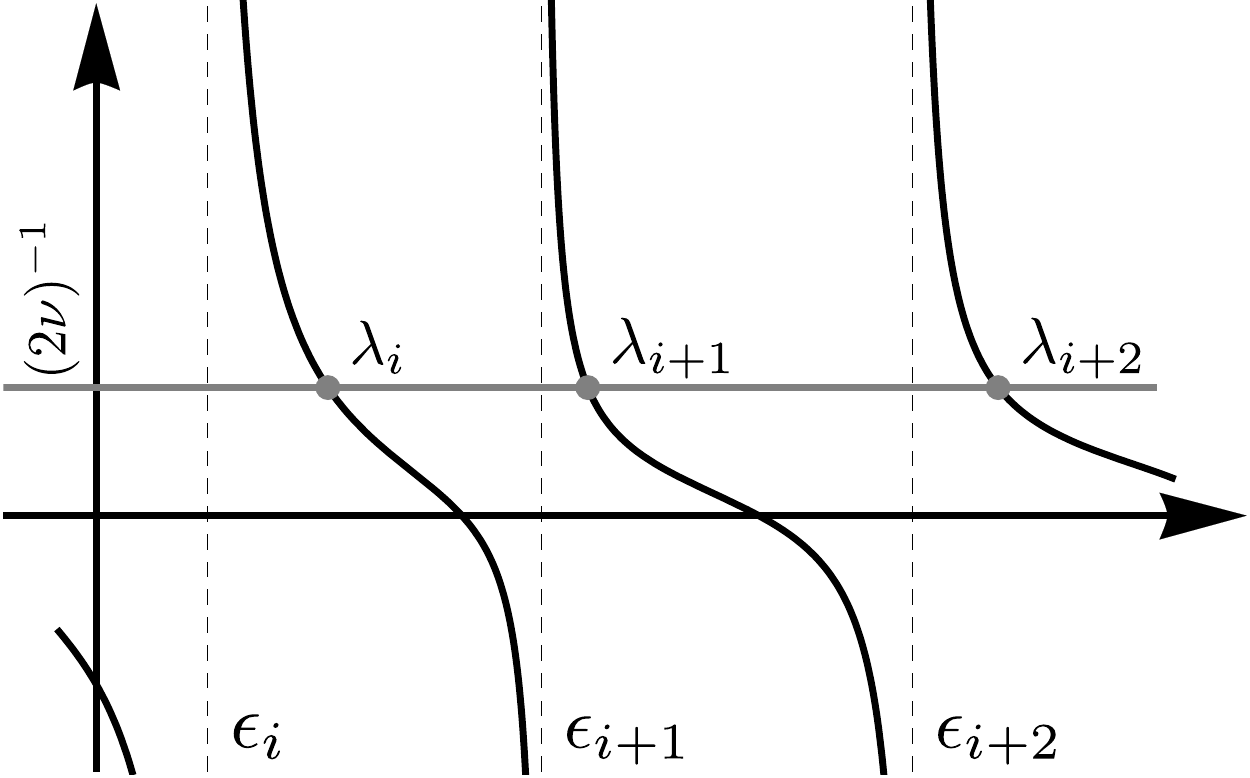}
\caption{Schematic representation of the right-hand side of eqs.~(\ref{eq:secEqGraphImag}, \ref{eq:secEqGraphSimp}) as a function of \(\lamb\). One finds a solution to the secular equation whenever this sum is equal to \( (2\nu)^{-1}\) , \(\nu\teq\tan{\alpha}\) as indicated by the horizontal, grey line. This graphical procedure nicely illustrates the interlacing property as all the \(\lamb_{i}\) can be found in between adjacent \(\eps_i\), \(\eps_{i+1}\) who define the poles of the sum. }
\label{pic:secGraph}
\end{figure}

At first the complex secular equation appears overdetermined to give \(B\) real solutions for \( \lamb \). This, however, is not true, since   the real part of this equation is trivially fulfilled due to the completeness condition \(\sum_{m=1}^B |A_m|^2\teq 1\), while the imaginary part yields
\begin{equation}
\cot{\alpha}=\sum_{m=1}^{B}\,|A_m|^2\cot{\left(\frac{\lamb-\eps_m}{2}\right)}\,.
\label{eq:secEqGraphImag}
\end{equation}
A prominent feature of   eq.~(\ref{eq:secEqGraphImag}) is the interlacing property satisfied by  its solutions, which is an immediate consequence of the fact that $S$ is a rank 1 perturbation of $S_0$.  Specifically, for a positive \( \alpha \) we  find that the following inequality holds, \( \epsilon_i \leq\lambda_i \leq\epsilon_{i+1} \). Similarly, in the case of \(\alpha<0\)  we have  \( \epsilon_{i-1} \leq\lambda_i \leq\epsilon_{i} \). This becomes apparent by noticing that the right hand side of equation~(\ref{eq:secEqGraphImag}), as a function of \(\lamb\), possesses poles at each \( \eps_m\).  Therefore, the solutions  \(\lamb_m\) lie in between those poles, see fig.~\ref{pic:secGraph}. In the limiting case of \( \alpha \to 0\) they will coincide with the poles as the system's spectrum is double degenerate.

\textbf{Large \(B\) limit.} Equation (\ref{eq:secEqGraphImag}) can be simplified further in the case of graphs with a large number of edges.  The eigenphases \(\eps_m\)  reside on the \( 2\pi \) interval parametrising the unit circle. Their distance therefore scales like \( O( \sfrac{1}{B}) \) with the number of edges.
Considering that \(\cot{x}\teq\sfrac{1}{x}\!+\!O(x)\), only the \(N \sim \sqrt{B}\) closest \(\eps_m\) will have an impact on the precise location of a solution \(\lamb_i\) between two \(\eps\)-eigenphases. To demonstrate  this  we split the sum in (\ref{eq:secEqGraphImag}) into three parts, the central sum with \(N\sim \sqrt{B}\) elements and two surrounding contributions with \(\sfrac{1}{2}(B-N)\) entries.  It is  straightforward to see  that due to the periodicity of the cotangent the contributions of the outer sums on average cancel each other.  Furthermore,  it can be shown that the fluctuations of those sums scale with \(O(\sfrac{1}{B})\) and therefore are negligible in the limit of large graphs.  After the expansion of the cotangent, we obtain for the remaining sum, up to corrections of \( O(\sfrac{1}{B})\):
\begin{equation}
\frac{1}{2\nu}=\sum_{m=1}^{N}
\frac{|A_m|^2}{\lamb-\eps_m}
\qquad\text{with}\quad
\nu=\tan{\alpha}\,.
\label{eq:secEqGraphSimp}
\end{equation}
The solutions of this equation obey the same interlacing property as before in eq.~(\ref{eq:secEqGraphImag}).

Remarkably, eq.~(\ref{eq:secEqGraphSimp}) is  identical to a secular equation considered in \cite{aleiner}. This work is concerned with the study of relations between old \(\eps_m\) and new \(\lamb_m\)-eigenvalues in a perturbed  Hamiltonian system, \( H= H_0 + \nu N \ket{\psi}\bra{\psi} \), where  \( H_0 \) is drawn from a GUE ensemble of  \(N \!\times\!N \) random matrices. Leaving aside  differences in the physical  interpretation this allows us to build upon the results of \cite{aleiner} in the next section, where we consider the spectral statistics of \(\{\lamb_m\}\).  As we demonstrate later,  the spectral statistics of the graph's physical spectrum \(\{k_n\}\) determined by  eq.~(\ref{eq:graphScatEv}) can, in the large \(B\) limit, be obtained from these results by a simple rescaling.
\section{Analytic Calculations. RMT approach.}
\label{sec:analyt}

Throughout this section we derive the nearest neighbour distance distribution \( \pnu(s) \) for nearly unidirectional quantum graphs using a RMT like approach.    In addition to the exact results presented here we provide a compact, heuristic surmise in section~\ref{sec:heuristic}.

Recall that the perturbed spectrum consists of two parts, the pinned \(\eps\)-eigen\-values, and in between the \(\lamb\) eigen\-values, \(\eps_i\le\lamb_i\le\eps_{i+1}\), moving under perturbation (see sec.~\ref{sec:secEQ}). Because of this it is  natural  to split \(\pnu(s) \) into two parts,
\begin{equation}
\pnu(s)=\frac{1}{2}\Big(\pex(s)+\pin(s)\Big)\,,
\end{equation}
where \( \pin(s) \) is the internal splitting distribution for the distances between \(\eps_i\) and the next \(\lamb_i \)  to the right while \( \pex(s) \) covers the expanse from \( \lamb_i\) to the next \(\eps_{i+1} \). The derivation of both splitting distributions follows along the same lines but we focus mainly on \( \pin(s) \) presenting results for \( \pex(s) \) at the end of this section. In addition, only the case \(\nu>0\) has to be treated.  As can be inferred from the secular equation,  the result  for \(\nu<0\) is obtainable by  exchanging the roles of \(\pin(s)\) and \(\pex(s)\), \ie
\begin{equation}
\pex(s)=p_{-\nu}^\text{in}(s)
\quad\text{and}\quad
\pin(s)=p_{-\nu}^\text{ex}(s)\,.
\label{ex:pRelationMnu}
\end{equation}
Let us emphasize that the results we present below for \(\nu>0\)  would not yield meaningful splitting distribution if negative \(\nu\) are entered, instead one should use the above relations.

Relating the splitting distributions \( p(s) \) to the gap probability  is a well established procedure in RMT since the later quantity is often easier to calculate, \cf \cite{haake}.
In our case we have two different sub-spectra and the gap probability  \( E \teq E(\emin,\emax; \lmin,\lmax) \) is defined as probability that  no eigenvalue of the respective kind, \ie \(\eps_i\) and \(\lamb_j\), can be found in the intervals \([\emin,\emax]\) and \([\lmin,\lmax]\), where, in principle, the two  intervals should be thought of as being independent.

To establish  the connection between   \( E \) and \( \pnu(s) \) let us first consider the probability   to find an \( \eps \)-eigenvalue in an interval \([\emin-\dte,\emin]\), in the limit of \( \dte\to 0 \). This  is the probability not to have an \(\eps\) gap at this position and thus can be expressed as \(1-E( \emin-\dte,\emin;0,0) \). Divided by the length \(\dte\) of the interval the result is the mean level density \(\bar{\rho}^{(\epsilon)}\) for the \(\eps\) spectrum

\begin{equation}
\bar{\rho}^{(\epsilon)}(\emin)=\frac{\partial}{\partial \emin}E(\emin,\emax;0,0)
\Big|_{\emin=\emax}\,. \label{eq:mlsDist}
\end{equation}

\begin{figure}[hbtp]
\centering
\includegraphics[width=0.55\textwidth]{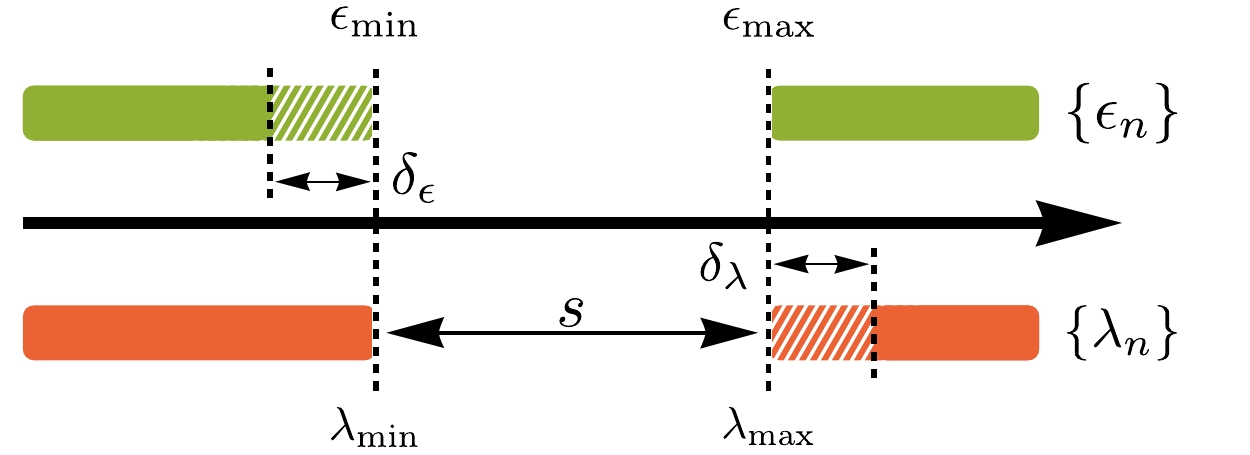}
\caption{\textit{(Colour online)} Drawing of a gap in the \(\{\eps_n\}\),\(\{\lamb_n\}\) spectrum with size \(s\). To take derivatives of the gap probability \(E\) we expand it linearly around small perturbations of the respective gap boundary, which one can think of as small \(\delta\) extensions of the original gap, see shaded intervals in the figure.}
\label{pic:gapSpec}
\end{figure}
We   now extend our consideration  to the gap intervals \([\emin,\emax]\),  \([\lmin,\lmax]\) where \(\emin\teq \lmin\),  \(\emax \teq \lmax\) while  the endpoints are separated by some finite distance  \(s\teq \bar{\rho}^{(\epsilon)}(\emin)|\lmin-\emax| \) measured on the scales of MLS \(\bar{\rho}^{-1}\), see fig.~\ref{pic:gapSpec}. Let  \(\plr\) be the 
 probability (density)  to find an \(\eps\)-eigenvalue in  the interval  \([\emin-\dte,\emin]\) simultaneously with  another \(\lamb\)-eigenvalue  in  the interval  \([\lmax,\lmax+\dtl]\) such that no other eigenvalues are present in-between. Following the same line of reasoning as before,
 it is given by a double derivative of  \( E \)  with respect to both edges \(\emin\) and \(\lmax\). On the other hand,  \(\plr\) is also equal to the probability density  \(\bar{\rho}^{(\epsilon)}\)  to find an \(\eps\)-eigenvalue  in the interval  \([\emin-\dte,\emin]\) times the probability  density to find the next \(\lamb\)-eigenvalue in the distance \(s \).
  The later one  is exactly the sought \(\pin(s)\). Therefore we may express it as
\begin{equation}
\pin(s)=\frac{ 1}{ \bar{\rho}^{(\epsilon)}(\emin)}\frac{\partial^2}{\partial \emin \partial \lmax}
E(\emin,\emax; \lmin,\lmax)
\big|_{\substack{\emin=\lmin=-s/2 \bar{\rho}^{(\epsilon)}(0) \\  \emax=\lmax=+s/2 \bar{\rho}^{(\epsilon)}(0)}},
\label{eq:pinDerivForm}
\end{equation}
where, for the sake of convenience, we have chosen the gap symmetrically around zero. 

A similar result holds for  \(\pex(s)\). In principle only the boundary derivatives have to be exchanged,
\begin{equation}
\pex(s)=\frac{ 1}{ \bar{\rho}^{(\lambda)}(\lmin)}\frac{\partial^2}{\partial \xl\partial\ye}
E(\emin,\emax; \lmin,\lmax)
\big|_{\substack{\emin=\lmin=-s/2\bar{\rho}^{(\lambda)}(0) \\  \emax=\lmax=+s/2 \bar{\rho}^{(\lambda)}(0)}}
\,,
\label{eq:pexDerivForm}
\end{equation}
where \(\bar{\rho}^{(\lambda)}\) denotes the mean level density of the \(\lambda\)-spectrum. Note that in our case  \(\bar{\rho}^{(\lambda)} \teq \bar{\rho}^{(\epsilon)}\).

\subsection{Gap Probability \(E\)}
\label{sec:analyt:gap}

We now  present the derivation of the gap probability  \(E\) for nearly unidirectional quantum graphs employing  a RMT model. Our  calculations rest upon the secular equation  (\ref{eq:secEqGraphSimp}), where the \(N\) distinct \(\eps\)-eigenvalues
 are distributed in accordance with the Circular Unitary Ensemble (\textit{CUE}) describing the distribution of eigenvalues in TRI unidirectional quantum graphs. 
Further on, we assume, in accordance with \cite{aleiner,portThomas}, a Gaussian distribution for  the overlaps  \(|A_m|^2\): 
\begin{equation}
p(|A_m|^2)=
N \exp{\left(-N |A_m|^2\right)}\,,
\label{eq:overlapDistribution}
\end{equation}
which can be expected for large graphs.  The validity of this assumption is discussed towards the end of section \ref{sec:comp}.

This  RMT model was employed previously  in \cite{aleiner} to calculate the two-point correlation function between the \(\{\eps_i\}\) and \(\{\lamb_i\} \) spectra, to which end the
following  joint probability distribution \( P(\{\eps_i\},\{\lamb_j\}) \) for the total spectrum was  derived:
\begin{equation}
P(\{\epsilon_i\},\{\lambda_j\})\propto 
\left(\prod_{\substack{i,j=1 \\ i>j}}^{N}
4 \sin{\frac{\epsilon_i-\epsilon_j}{2}} 
\sin{\frac{\lambda_i-\lambda_j}{2}} \right)
\exp{\left(-\frac{N}{2\nu}\sum_{i=1}^{N}(\lambda_i-\epsilon_i)\right)}\,.
\label{eq:jointPDF}
\end{equation}
The  \( \eps \) and \( \lamb \) eigenvalues run  from \( [-\pi,+\pi] \) under the constraint of the interlacing property. Please observe that \(\{\eps_i\}\) is taken without degeneracy and \(\{\lamb_i\}\) only consists of the shifted eigenvalues, similar to the usage in the secular equations.
We will now utilize (\ref{eq:jointPDF}) in order to calculate the gap probability function: 
\begin{equation}
\begin{split}
E & (\xe,\ye;\xl,\yl)
\propto \ints{-\pi}{+\pi}{\eps_1}\ints{\eps_1}{\pi}{\lamb_1} \ints{\lamb_1}{\pi}{\eps_2} \ldots \ints{\eps_N}{+\pi}{\lamb_N}\,
P(\{\eps_i\},\{\lamb_j\})\\
&\times\prod_{k=1}^{N}\Big( \big(1-\theta(\eps_k-\xe)\theta(\ye-\eps_k)\big) \big(1-\theta(\lamb_k-\xl)\theta(\yl-\lamb_k)\big) \Big)\,.
\end{split}
\end{equation}
A special feature of the gap probability is that it relates to all eigenvalues of the spectrum in a uniform manner, visible in its product structure,
which  allows to take all integrals  by standard methods, \cf \cite{haake}. In the following we will briefly sketch this procedure.

Expanding the sine functions into exponentials,
\begin{equation}
\prod_{\substack{i,j=1\\i>j}}^{N}\sin{\left(\frac{\eps_i-\eps_j}{2}\right)}
\propto \prod_{\substack{i,j=1\\i>j}}^{N} \eu^{-\iu/2(\eps_i+\eps_j)}\left(\eu^{\iu\eps_i} -\eu^{\iu\eps_j}\right)
= \Delta{\{\eu^{\iu\eps_i}\}}\,\prod_{j=1}^{N}\eu^{-\frac{\iu}{2}(N-1) \eps_j}\,,
\label{eq:sinproddecomp}
\end{equation}
one finds Vandermonde determinants \( \Delta \) of  \( \eps \) variables with  a  similar result  for  the \(\lamb \) part.  Upon reordering the domains of integration,
\begin{equation}
\ints{-\pi}{+\pi}{\eps_1}\ints{\eps_1}{\pi}{\lamb_1} \ints{\lamb_1}{\pi}{\eps_2} \ldots \ints{\eps_N}{+\pi}{\lamb_N}
\rightarrow
\ints{-\pi}{+\pi}{\eps_1}\ints{\eps_1}{+\pi}{\eps_2}\ldots
\ints{\eps_{N-1}}{+\pi}{\eps_N}\ints{\eps_1}{\eps_2}{\lamb_1}
\ldots\ints{\eps_{N-1}}{\eps_N}{\lamb_{N-1}} \ints{\eps_N}{+\pi}{\lamb_N}\,,
\label{eq:reorderingIntDom}
\end{equation}
the  integrals  over the \(\lamb\) variables can be drawn into the corresponding determinant form:
\begin{equation}
D\{\eps_i\}=\left|
\begin{array}{cccc}
\ints{\eps_1}{\eps_2}{\lamb_1}f_1(\lamb_1) & \ints{\eps_1}{\eps_2}{\lamb_1}f_2(\lamb_1) & \ldots & \ints{\eps_1}{\eps_2}{\lamb_1}f_N(\lamb_1) \\
\vdots & \vdots & \ddots & \vdots \\
\ints{\eps_{N-1}}{\eps_N}{\lamb_{N-1}}f_1(\lamb_{N-1}) & \ints{\eps_{N-1}}{\eps_N}{\lamb_{N-1}}f_2(\lamb_{N-1}) & \ldots & \ints{\eps_{N-1}}{\eps_N}{\lamb_{N-1}}f_N(\lamb_{N-1}) \\
\ints{\eps_N}{\pi}{\lamb_N}f_1(\lamb_N) & \ints{\eps_N}{\pi}{\lamb_N}f_2(\lamb_N) & \ldots & \ints{\eps_N}{\pi}{\lamb_N} f_N(\lamb_N)
\end{array}
\right|\,,
\label{eq:symlamdetintpar1}
\end{equation}
where \(f_j(\lamb_i)\) is a shorthand for \( \exp{(\iu(j-1)\lamb_i +\iu(N-1)\lamb_i /2 -N\lamb_i/(2\nu))} \).
The remaining integral over the \(\eps\) variables
\begin{equation}
E  (\xe,\ye;\xl,\yl)\propto\ints{-\pi}{+\pi}{\eps_1}\ints{\eps_1}{+\pi}{\eps_2}\ldots
\ints{\eps_{N-1}}{+\pi}{\eps_N}D\{\eps_i\}\Delta{\{\eu^{\iu\eps_i}\}}\prod_{j=1}^{N}\eu^{\frac{\iu N\eps_j}{2}\left(1-\frac{1}{N}- \frac{\iu}{\nu}\right) }\,,
\label{eq:symlamdetintpar3}
\end{equation}
can then be treated as follows.
Adding the last line of the determinant \(D\{\eps_i\}\) to the second last and then continuing recursively sets the upper boundary to \(\pi\) in all integrals and shows the antisymmetry of \(D\{\eps_i\}\)  under exchange of \(\eps_i \leftrightarrow \eps_j\). Therefore, the  integrand in eq.~(\ref{eq:symlamdetintpar3}) is a  symmetric function, implying that  the boundaries of integration can be extended to the full domain \([-\pi,+\pi]\) (up to a factor). 

Finally,  we exploit the fact that Vandermonde determinants can be expanded into alternating sums
\begin{equation}
\Delta{\{\eu^{\iu\eps_i}\}}
=\sum_{\{\sigma\}}(-1)^{|\sigma|}\eu^{\iu \sigma(1) \eps_1}
\eu^{\iu \sigma(2) \eps_2}\ldots\eu^{\iu \sigma(N) \eps_N}\,,
\label{eq:detExpVan}
\end{equation}
wherein \(\{\sigma\}\) represents the set of all permutations of the numbers \((0,\;1,\;\ldots\;N-1)\), \(|\sigma|\) denotes the permutation's parity and \(\sigma(i)\) yields the number associated to \(i\) under permutation \(\sigma\).
The symmetry of the integrand in (\ref{eq:symlamdetintpar3}) allows us to  absorb the single factors into the second determinant, leading to
\begin{equation}
E(\xe,\ye;\xl,\yl)= \frac{\det F(\xe,\ye;\xl,\yl)}{\det F(0,0;0,0)}\,,
\label{eq:EasDet}
\end{equation}
where \(F(\xe,\ye;\xl,\yl)\) is a \( N\times N\) matrix  with elements
\begin{equation}
\begin{split}
F_{kl}=&\ints{-\pi}{+\pi}{\eps}\ints{\eps}{+\pi}{\lamb}\,
\eu^{-\frac{N}{2\nu}(\lamb-\eps)}
\eu^{\iu(k-1)\eps-\iu\frac{N-1}{2}\eps}
\eu^{\iu(l-1)\lamb-\iu\frac{N-1}{2}\lamb}
\\
&\times\Big( \big(1-\theta(\eps-\xe)\theta(\ye-\eps)\big) \big(1-\theta(\lamb-\xl)\theta(\yl-\lamb)\big) \Big)\,.
\end{split}
\label{eq:fklInt}
\end{equation}
The denominator of \(E\) ensures that the probability to find gaps of zero width in both spectra is unity. It is related to the omitted normalisation of \(P\)  in Equation (\ref{eq:jointPDF}).
The integral in (\ref{eq:fklInt}) consists of four parts, depending on the combinations of Heaviside \( \theta \)-functions, which can be calculated explicitly. For the sake of compactness of exposition the resulting expressions are not presented  here. They can be found in the appendix.

\subsection{Splitting Distribution}
\label{sec:analyt:split}
Using eqs.~(\ref{eq:pinDerivForm}, \ref{eq:EasDet}) and  taking  into  account that \(\bar{\rho}^{(\lambda)}\teq \bar{\rho}^{(\epsilon)}\teq N/(2\pi) \) we can write the nearest neighbour distribution as  
\begin{equation} \pin(s)=
\left(\frac{2\pi}{N}\right)\frac{\partial_{\dte} \partial_{\dtl} \det F^\text{(num)} (s\pi/N,\dte,\dtl) }{\partial_{\dte}  \det F^\text{(den)} (s\pi/N,\dte)} \Big{\vert}_{\dte=\dtl=0}, \label{eq:InDistr}
\end{equation}
where the matrices in the  numerator and denominator are given by
\begin{align}
F^\text{(num)}(z,\dte,\dtl) & :=
F(-z-\dte,
z;-z,
z+\dtl)\,,
\label{eq:NochZweiMatr1}\\
F^\text{(den)}(z,\dte) & :=
F(-z-\dte,-z;0,0).
\label{eq:NochZweiMatr2}
\end{align}
For an illustration of the gap position please refer to figure \ref{pic:gapSpec}.
Expression (\ref{eq:InDistr}) can be straightforwardly evaluated expanding \(F^\text{(num)}\) and \(F^\text{(den)}\)
 up to linear order in \(\dte \cdot\dtl \)  (resp.  \(\dte  \)) and then taking the  large  \(N\)-limit. (The resulting expressions are given by equations (\ref{eq:fklNumerator}) and (\ref{eq:fklDenominator}) in the appendix.) At this point it is convenient to introduce  
a pair of auxiliary  \(N\)-dimensional vectors  \(\ket{u}, \ket{u^{*}}\):
\begin{equation}
\langle j\ket{u}= \eu^{+\iu j \frac{\pi}{N} s} \,,
\quad
\langle j\ket{u^{*}}= \eu^{-\iu j \frac{\pi}{N} s} \,, \label{eq:lambdaDef}
\end{equation}
and  \(N\times N\) matrices:
\begin{equation}
\Lambda_{kl}= \left(\iu k+\spert-\iu\frac{N+1}{2}\right)\delta_{kl} \,,  \qquad 
R_{kl}=\delta_{kl}-\left(\frac{\sin{(k-l)\sfrac{\pi}{N}s}}{(k-l)\pi}\right)_{kl}\,.
\label{eq:defR}
\end{equation}
Note that   \( 1-R \) is the well known sine-kernel matrix. 
This  notation enables us to write the leading order expansion of  both matrices (\ref{eq:NochZweiMatr1}, \ref{eq:NochZweiMatr2}) in a compact form:
\begin{equation}
F^\text{(den)}=-2\pi \Lambda
+\dte\Lambda\ket{u}\bra{u}+O(\sfrac{1}{N})\,,
\label{eq:fDenom}
\end{equation}
\begin{equation}
F^\text{(num)}=
-2\pi\Lambda R+\ket{u}\bra{u}
-g \big(\dte\Lambda\ket{u^{*}}-\ket{u^{*}}\big)\,
\big(\bra{u}\Lambda\dtl-\bra{u}\big)+O(\sfrac{1}{N})\,,
\label{eq:fNum}
\end{equation}
where \(g=\exp{\left( \iu(N+1)\frac{\pi}{N}s -\frac{\pi}{\nu}s \right)}\). The corrections  stand here for   neglected terms of order  \(O(\sfrac{1}{N})\)  in the elements of  \(F\).
Employing  the relationship
\begin{equation}
\det(A+\ket{x}\bra{y})=
\det(A)\left(1+\bra{y}A^{-1}\ket{x}\right()
\end{equation}
the  determinant of \(F^\text{(den)}\) can easily be calculated:
\begin{equation}
\det(F^\text{(den)}) =
(-2\pi)^N \det(\Lambda)(1-{N\dte}/{2\pi})\,.
\end{equation} 
Taking also into account
\begin{equation}
\big(A+\ket{u}\bra{v}\big)^{-1}=
A^{-1}-\frac{A^{-1}\ket{u}\bra{v}A^{-1}}{1+\bra{v}A^{-1}\ket{u}}\,,
\end{equation}
 we can perform a similar expansion for the  determinant of \(F^\text{(num)}\), leading finally to
\begin{equation}
\begin{split}
\pin(s)= 2\frac{g\pi}{N^2}\det{R}
\times \Big( & \bra{u} R^{-1} \Lambda \ket{u^{*}}
-\frac{1}{2\pi}\bra{u} \Lambda^{-1} R^{-1} \ket{u}\bra{u} R^{-1} \Lambda \ket{u^{*}} \\
& +\frac{1}{2\pi}\bra{u}R^{-1}\ket{u} \bra{u}\Lambda^{-1} R^{-1} \Lambda \ket{u^{*}} \Big)
+O(\sfrac{1}{N})\,.
\end{split}
\label{eq:pinFinal}
\end{equation}
Note that, although   this quantity  explicitly depends on \(N\), it has a well defined limit for \(N\to\infty \).  The leading denominator is compensated by the corresponding  scaling   of the scalar products. Furthermore, due to the presence of the exponent $g$ the above expression turns out to be  purely  real. 

Along the same lines we are able to obtain the result for \(\pex(s)\) using eq.~(\ref{eq:pexDerivForm}).  
Expanding the double derivative  up to the leading order   in \(\dte, \dtl\) yields:
\begin{equation}
\pex(s)=\frac{\det{R}}{N^2}
\Big(
\braket{u^{*}|R^{-1}|u^{*}}
\braket{u|\Lambda^{-1} R^{-1} \Lambda | u}
-
\braket{u|\Lambda^{-1} R^{-1} | u^{*}}
\braket{u^{*} | R^{-1} \Lambda | u}
\Big)+O(\sfrac{1}{N})\,.
\label{eq:pexFinal}
\end{equation}
It is interesting to compare this result with the GUE nearest neighbour  distribution \(\pgue(s)\)  which should emerge from (\ref{eq:pexFinal})
in the limit of \(\nu\to 0\). To this end note that   \(\pgue(s)\) is related to the \(R\) matrix, \cite{haake}, as:
\begin{equation}
\pgue(s)=\partial^2_s \det{R}
=\det{R}\times
\left( \partial^2_s\Tr\log{R}
+(\partial_s\Tr\log{R})^2\right)\,.
\end{equation}
Further on, the derivative of \(R\) can be expressed as a rank 2 projector on \(\ket{u}\) and \(\ket{u^*}\). In particular
\begin{equation}
\partial_s\Tr\log{R}=\Tr{\frac{\partial_s R}{R}}
=-\frac{1}{2N}\left(
\bra{u}R^{-1}\ket{u}+\bra{u^{*}}R^{-1}\ket{u^{*}}
\right)\,.
\end{equation}
The second derivative can be treated in the same fashion and, after some cancellation, we obtain
\begin{equation}
\pgue(s)=\frac{\det{R}}{N^2}
\Big(
\braket{u^{*}|R^{-1}|u^{*}}
\braket{u| R^{-1}  | u}
-
\braket{u| R^{-1} | u^{*}}
\braket{u^{*} | R^{-1} | u}
\Big)\,,
\label{eq:pgueFinal}
\end{equation}
which bears an apparent  structural similarity to (\ref{eq:pexFinal}). 
\section{Comparison with Quantum Graph Spectra}
\label{sec:comp}

Both  expressions (\ref{eq:pinFinal},\ref{eq:pexFinal})  depend on the inverse of the sine-kernel matrix  \(R^{-1}\) which is not known explicitly. Nevertheless, they serve as a very useful tool to numerically calculate the nearest neighbour distribution.  As we neglected terms of the order 
\(O(1/N) \),  the  dimension \( N \) of the matrix \(R\) should be large
enough to  reach the limiting distribution with a sufficient precision.  For practical purposes we found
\(N\teq 100 \) to be sufficient  for all ranges of the parameter strength \(\nu\).

\begin{figure}[hbtp]
\centering
\includegraphics[width=0.85\textwidth]{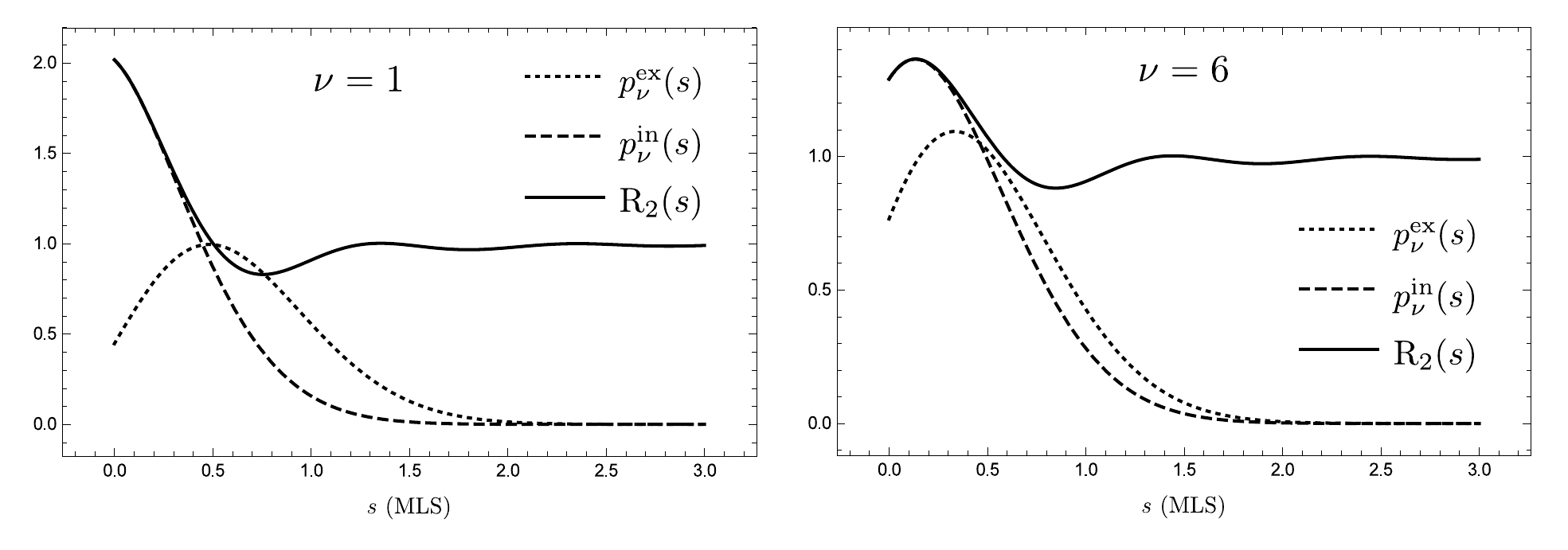}
\caption{Analytic results for \(\pin(s)\), \(\pex(s)\) and the \(R_2(s)\) two point correlator derived
by \cite{aleiner} shown together for \(\nu\teq 2\) and \(\nu\teq 6\). As expected \(\pin(s)\) and \(R_2(s)\) are almost
identical for small \(s\). But for larger values \(\pin(s)\) decays to 0 as the
nearest neighbour can not be arbitrarily far, while \(R_2(s)\) saturates towards 1 as
farther away values are no longer correlated. With increasing perturbation strength, see right panel, \(\pin(s)\) and \(\pex(s)\) tend towards the same limiting distribution.}
\label{pic:r2comp}
\end{figure}

The resulting nearest neighbour distributions obtained evaluating eqs.~(\ref{eq:pinFinal},\ref{eq:pexFinal}) are presented  in fig.~\ref{pic:r2comp}. In all cases we found a perfect agreement with the distribution (not shown here) of eigenvalues drawn from the Random Matrix Ensembles of sec.~\ref{sec:analyt}. Without perturbation, \ie for \(\nu\teq 0\),  \(\pex(s) \)  is  given by the  GUE nearest neighbour 
distribution while  \(\pin(s)\)   is  just a \(\delta\)-spike at 0 due to the exact degeneracy of
the system. Under a small perturbation this spike erodes, but both parts of the splitting
distribution, \(\pin(s)\) and \(\pex(s) \), are still distinguishable. With increasing
perturbation strength both distributions become more similar. From the secular equation
(\ref{eq:secEqGraphSimp}) one can  infer that, in the limit \(\nu\to\infty\), they  are actually   identical.

 It is quite remarkable that the perturbation does not
lead to a strict level repulsion as for any \(\nu\) one finds \(\pin(0)>0\).
Furthermore, under onset of the perturbation there is no longer a strict repulsion
between \(\lamb_i\) and the \(\eps_{i+1}\) eigenvalue to the right either, \ie
\(\pex(0)>0\) if \(\nu>0\).
For comparison we present here the  two point correlation function \(R_2(s)\)
between the \(\eps\) and the \(\lamb\) part of the spectrum which was derived in \cite{aleiner, simon} (for related quantities also refer to \cite{hentschel}). For small distances \(s\) this function naturally agrees with \(\pin(s)\) as is visible in fig.~\ref{pic:r2comp}.

\textbf{ Nearly unidirectional quantum graphs.}
In what follows we compare the above results 
with the   nearest neighbour
distributions of actual quantum graph spectra \(\{k_n\}\). For this we use several families of
nearly unidirectional quantum graphs constructed according to the
guidelines in  sec.~\ref{sec:qg1}.  To find numerical  solutions to the equation
\begin{equation}
\det \big(1\!-\!S\mathcal{L}(k_n)\big)=0 
\end{equation} 
 we used the (by definition) positive singular eigenvalues of \(1\!-\!S\mathcal{L}(k) \) and searched for the points where the lowest one becomes 0, see \eg \cite{Baecker} for details of this method.
Noticing that, for not too small graphs, the lowest lying eigenvalues depend  approximately linearly on  \( k\)   it is straightforward to follow the
downwards slope within a few iterations. We found this
approach to work rather fast and accurately, such that 9 significant post-decimal digits
 can easily be achieved.

In addition to the  actual \(\{k_n\}\) spectrum  of the graph we also computed  the splitting distribution between 
the eigenphases of the unitary quantum map \( S\mathcal{L}(k) \).  In the large \(B \) limit both spectra are  known to possess  the same  statistics after  rescaling the MLS to \(1\), \cite{graphsReview,berkolaiko}. Indeed, the average
``velocity'' of the eigenphases over \(k\) depends on the (fixed) average edge length \(\bar{l_i}\) while fluctuations
decrease with increasing bond number. As a result, both spectral statistics
coincide in the large \(B\) limit. Note, however, that since we obtain
\(2B \) eigenvalues for any arbitrary \(k\), this approach is numerically significantly less expensive.

If not stated otherwise we set the
back-scattering matrix \(\tilde{\sigma}\), eq.~(\ref{eq:backscatSigma}), to
\begin{equation}
\begin{split}
\alpha=-\arctan{\nu}\,,\quad
\beta=0\,,\quad
\gamma=0 
\\
\rightarrow\qquad
r=\frac{\nu}{\iu-\nu}\,,
\quad
t=\frac{\iu}{\iu-\nu}\,.
\end{split}
\label{eq:choiceABG}
\end{equation}
This choice is reminiscent of placing an actual strength \(\nu\) \(\delta\)-potential on an edge ( \(H=\Delta+\nu\delta(x) \)), where transmission
and reflection rates of the free wave propagation are given by \(r\) and \(t\).
We omit here the \(k\) dependence of such type of perturbation.  The graph lengths were chosen randomly from the interval \([0,1] \) and rescaled such that the mean length is \(1\).

\begin{figure}[hbtp]
\centering
\includegraphics[width=0.9\textwidth]{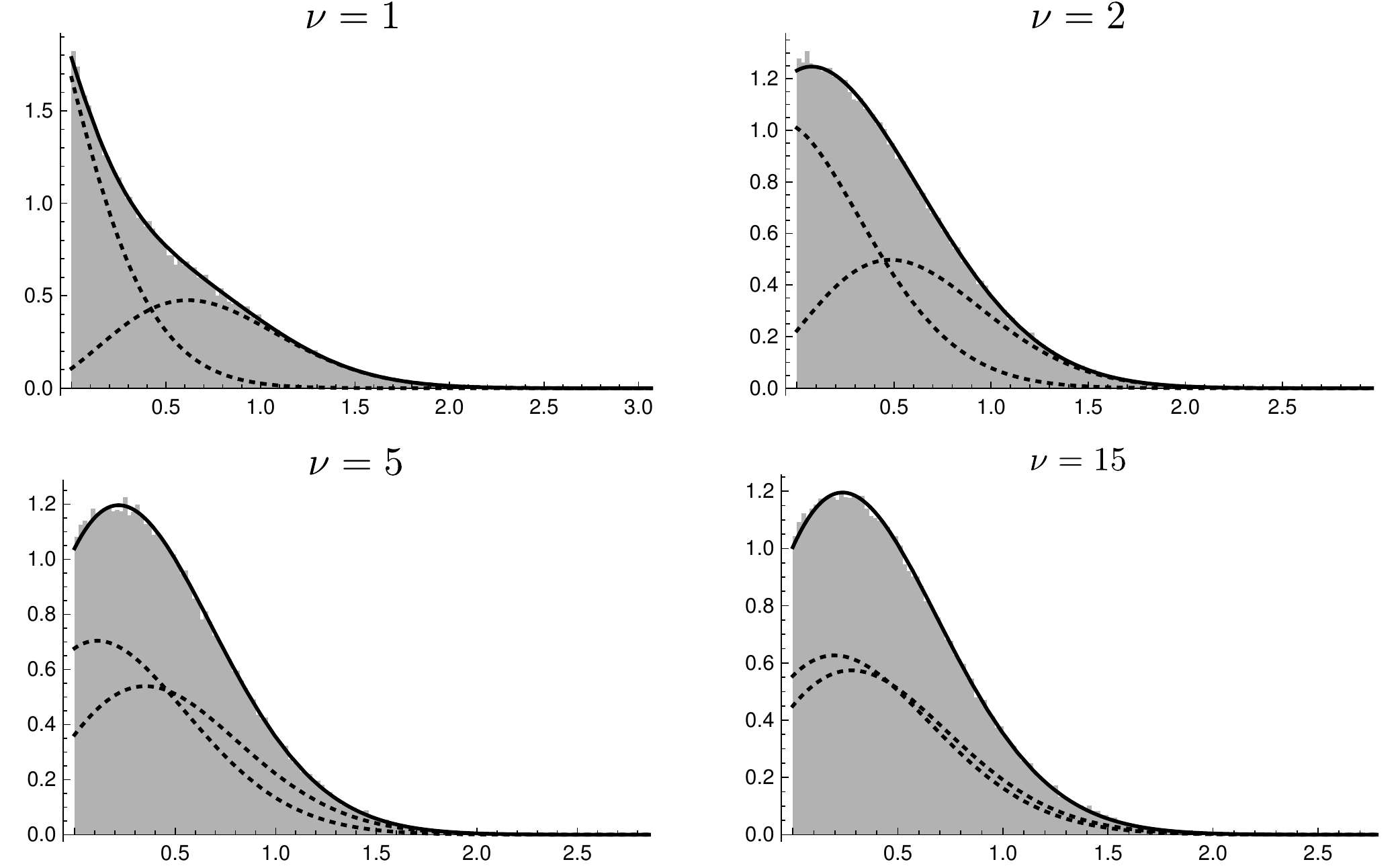}
\caption{Histograms of the nearest neighbour distances for the 130,000 lowest
\(k\)-eigenvalues obtained from a fully connected graph with 17 vertices plus
back-scatterer, see text. The black line shows the analytic result derived in
sec.~\ref{sec:analyt} for comparison. The smaller dashed lines indicate the
contributions of the single \(\pin(s)\) and \(\pex(s)\) to the analytical result. To
distinguish them please note that  \(\pin(0)\geq\pex(0)\). The
lower panels demonstrate the onset of saturation for strong perturbations in which
the distribution of both sub-splittings will become identical. MLS is adjusted to \(
\braket{ \eps_{i+1} -\eps_i}=1 \). (Weaker perturbation strengths can be found in
fig.~\ref{pic:fullyKscat})}
\label{pic:fullyK}
\end{figure}

\begin{figure}[hbtp]
\centering
\includegraphics[width=1.\textwidth]{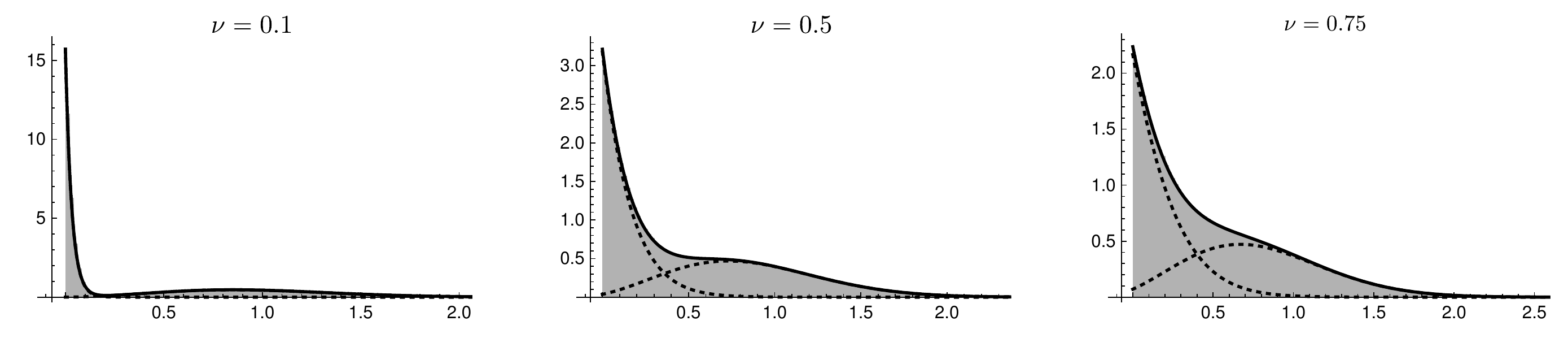}
\caption{Histograms for \(\pnu(s)\) obtained from \( 2.7\!\times\! 10^6\)
eigenphases of the graph's quantum map. Further details as in fig.~\ref{pic:fullyK}.}
\label{pic:fullyKscat}
\end{figure}

\textit{Fully connected graph.} This  graph is composed of  \(V\)  vertices, which are interconnected by \(B=V(V-1)/2\) bonds. Note that we do not allow self-loops, \ie a vertex cannot be connected to itself.   
 At each vertex the \(U_i\)  scattering matrices, see eq.~(\ref{eq:locScatForm}), are drawn randomly. Figs.~\ref{pic:fullyK} and \ref{pic:fullyKscat}  demonstrate
that both \(\pin(s)\) and \(\pex(s) \)  agree quite nicely with the analytic predictions.

\textit{Binary De Bruijn Graphs.} The \(2^p\)  vertices of these graphs can be labelled by binary sequences of  length \(p\). Each  vertex \(a_1 a_2\dots a_p\), \(a_i=\{0,1\}\) is connected  with several (generically \(4\)) others labelled by the sequences \(b a_1 a_2\dots a_{p-1}\),  \(b=\{0,1\}\),  or \( a_2 a_3\dots a_p c\),  \(c=\{0,1\}\), which are obtained by adding one symbol to the
left and removing one on the right or vice versa, see \cite{boris3}. A sketch of such a graph for  \(p\teq 3\) can be found in fig.~\ref{pic:unidirecGraphSchema}. 
As opposed to the fully connected graphs, De Bruijn graphs feature several short cycles -- the alternating pattern \(1010\ldots 0\to 0101\ldots 1 \to
1010\ldots 0 \) is of length 2, while the vertices \( 11\ldots 1\) and \(00\ldots 0\) have
attached self-loops.  The scattering matrices at each vertex are identical and chosen in such a way that the original graph is unidirectional. 

\begin{figure}[hbtp]
\centering
\includegraphics[width=0.9\textwidth]{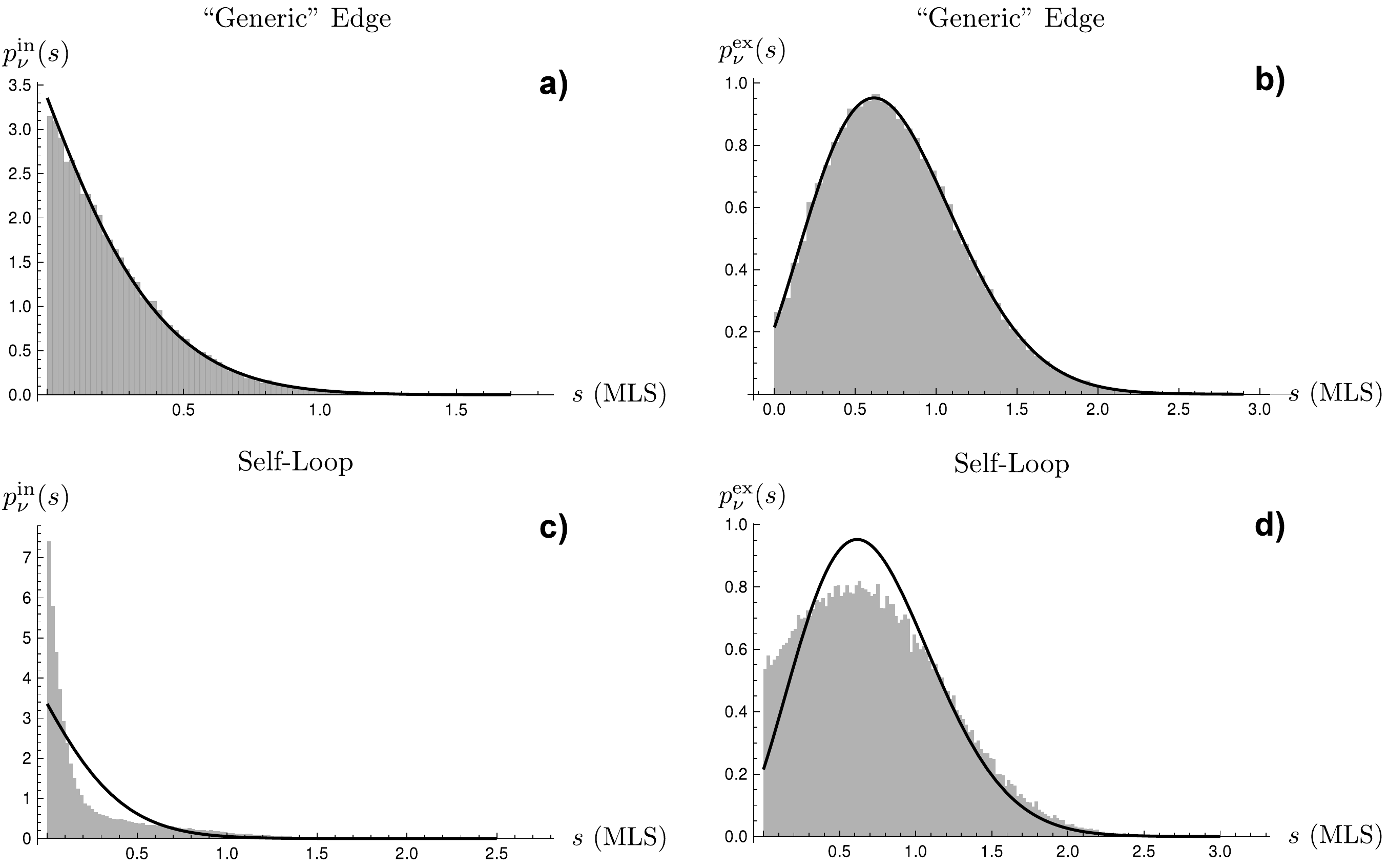}
\caption{Histogram for the splitting distributions in binary De Bruijn graphs. In
the lower row the back-scattering element is located on a self-loop of the graph
while in the upper it is far away from short cycles. It is plainly visible that the ``self-loop'' 
distributions deviate largely from the anticipated analytic results (black lines)
while for the ``generic'' case the agreement is comparable to the results obtained
from the fully connected graph, figs.~\ref{pic:fullyK} and \ref{pic:fullyKscat}.
Further on, it appears that the effect is strongest for small splittings \(s\) and
therefore mainly affects the internal splitting \(\pin(s)\).}
\label{pic:brokenDeBruijn}
\end{figure}

\begin{figure}[hbtp]
\centering
\includegraphics[width=0.9\textwidth]{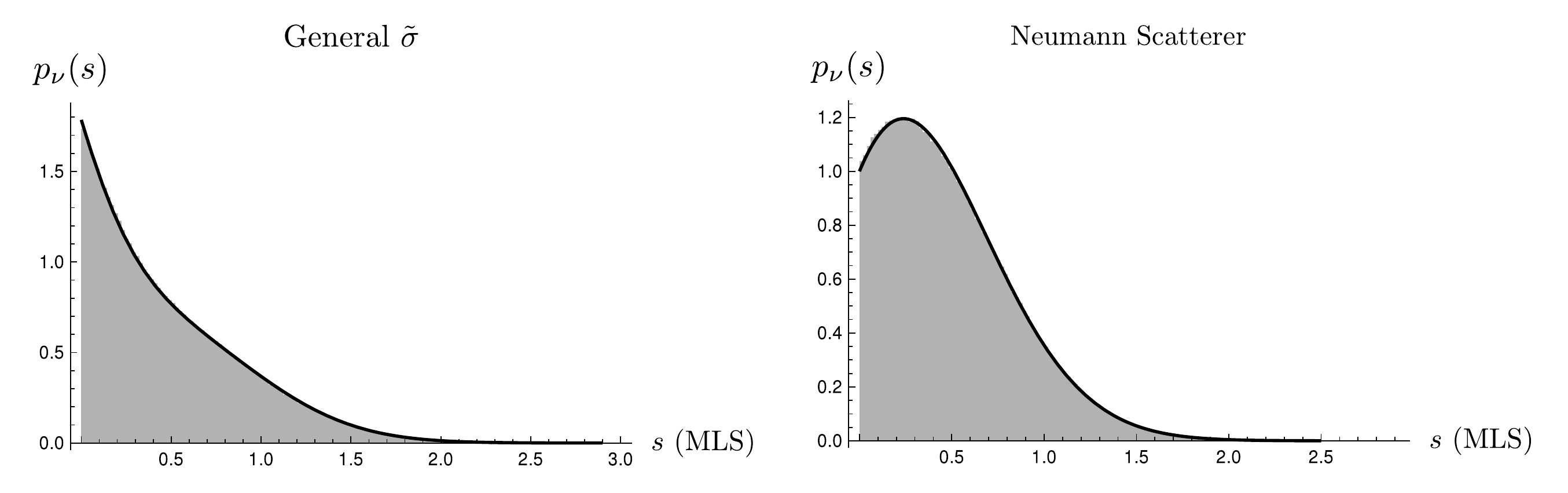}
\caption{Depicted are nearest neighbour distance distributions for a De
Bruijn graph with 64 vertices. On the left-hand side we placed, on a ``generic'' edge (see fig.~\ref{pic:anStat}), a general scatterer
with \(\alpha\teq \pi/4\), \( \beta\teq \pi/3\) and
\(\gamma\teq 4/5\,\pi\), which corresponds to \(\nu\teq 1\) as indicated by the
analytic black line.
On the right-hand side no scatterer is placed at all, as a substitute we replace \(\sigma_i\) of one vertex with Neumann boundary conditions. The plotted analytical
result represents \(\nu\!\to\!\infty\). Minor deviations can be
attributed to the mismatch in \(p(|A_m|^2)\) for De Bruijn graphs, see right-hand
side of fig.~\ref{pic:anStat}}
\label{pic:otherScat}
\end{figure}

The results for the nearest neighbour distributions  turn out to be   quite sensitive to the choice  of 
the back-scatterer position. If  \(\tilde{\sigma}\) is placed on a generic edge  we once again find a good agreement with the RMT predictions, see fig.~\ref{pic:brokenDeBruijn}a,b. As can be seen on fig.~\ref{pic:otherScat}, the same stays true if we change
 the  scattering matrix  at some generic vertex of  the original unidirectional quantum graph \(\Gamma_0\) to correspond to Neumann boundary conditions (thus breaking unidirectionality).   On the other hand, if we place the back-scatterer on one of the self-loops (see  fig.~\ref{pic:unidirecGraphSchema}) this has a drastic effect on the resulting splitting distribution, see fig.~\ref{pic:brokenDeBruijn}c,d.
 
\begin{figure}[hbtp]
\centering
\includegraphics[width=0.9\textwidth]{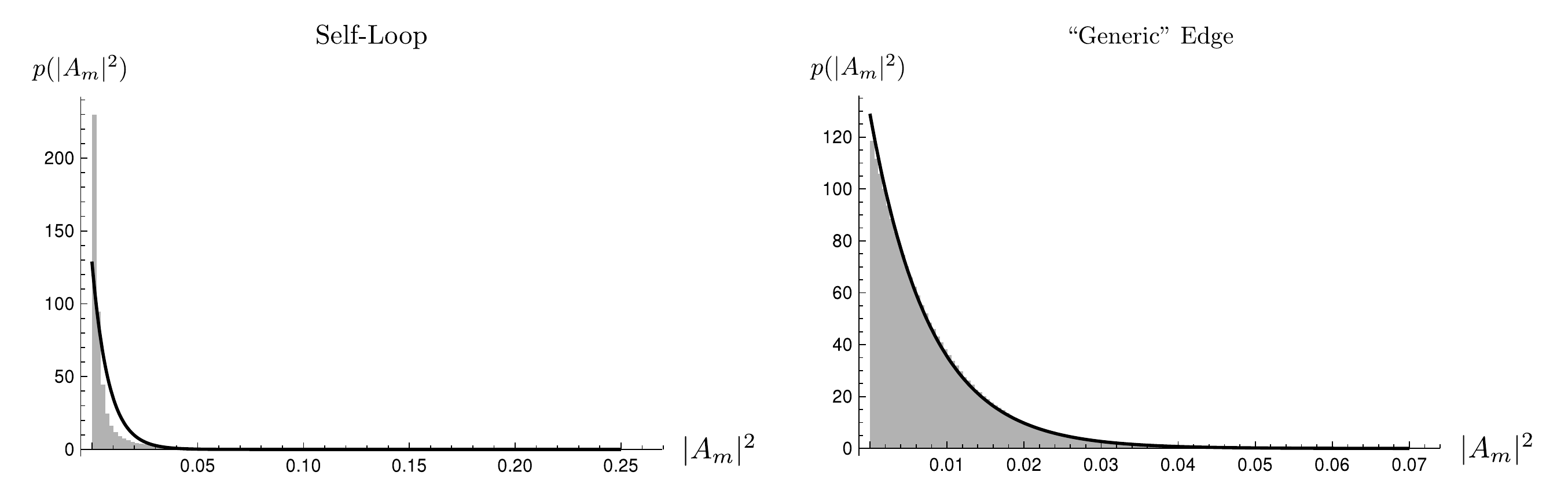}
\caption{Histograms showing the distribution of wave functions (absolute square) on
two edges of a De Bruijn graph. Either on a self-loop or far away from any short
cycles (right panel). In the generic case one finds the expected uniform
distribution as in eq.~(\ref{eq:overlapDistribution}). But for the self-loop
the wave function either avoids the edge, \ie increased probability for very small
\(A_m\) or it has an enhanced localisation. A logarithmic scaling would reveal an (exponentially decaying) heavy-tail for large \(A_m\) where the distribution is orders of magnitudes larger
than our assumption suggested. For instance \(|A_m|^2>0.2\) denotes that at least
one fifth of the wave-function is localised on this edge, implying a much smaller weight on the other 127 edges.}
\label{pic:anStat}
\end{figure}

To explain these findings let us recall that in the
analytic RMT model  we made an assumption, eq.~(\ref{eq:overlapDistribution}), on the uniform (random wave) distribution of the wave-function's
probability density \(|A_m|^2\) at the scatterer position. On the other hand,  wave functions  on certain types of graphs  are known to exhibit enhanced localisation (scars) on some edges \cite{scars,notDaniel}. To shed further light onto the sensitivity of \(\pnu(s)\) to such enhancements
we analysed the distribution of  \(|A_m|^2\) at different edges of the graph. Fig.~\ref{pic:anStat}
shows  the results  for two different edges of  a De Bruijn graph illustrating
the significant differences between the actual result and our original assumption,
eq.~(\ref{eq:overlapDistribution}).

\textit{Other systems.} Besides the case of fixed local scattering matrices \(\sigma_i\) we also investigated the case of a random choice for the De Bruijn graphs, as well as completely randomly constructed unidirectional graphs with low connectivity. Above deviation was present in all short cycles (containing up to 4 edges) of the tested graphs, decaying with increasing cycle length. Notwithstanding, the choice of the local scattering matrices \(\sigma_i\) along the cycles also has an important impact on the strength of the effect.
\section{Heuristic Surmise}
\label{sec:heuristic}

The  formulas (\ref{eq:pinFinal},\ref{eq:pexFinal}) for the nearest neighbour distribution, although  exact,  require to calculate the inverses of large matrices.  It would therefore be
of interest  to have a simple, analytical expression approximating    \(\pin(s)\) and  \( \pex(s) \).  In the following we obtain such an expression based on the Wigner Surmise providing the nearest neighbour distribution for GUE:
 \begin{equation}
\pw(s)=
32\, \frac{s^2}{\pi^2} \exp{\left(-\frac{4 s^2}{\pi }\right)}\,.
\label{eq:wignerSurmise}
\end{equation}  

First observe  that independently of the perturbation strength \( \nu \),  the \( \eps \) and the shifted \( \lamb \) part of the spectrum are both GUE distributed  if considered separately, \cite{bogomolnyScatterer}. Yet, these distributions are not independent, since  we have the  interlacing property -- a new eigenvalue \( \lamb_{i} \) is at least as far from \( \lamb_{i-1} \) as from \( \eps_{i} \).  To take into  account  correlations between the \( \{\eps_i\}\)  and \(\{\lamb_i\}\) spectra let us  make a crude assumption that \( \eps_{i} \) and  \( \lamb_{i-1} \) are separated by a fixed distance  
\(\cin \). Since the distances between \( \lamb_{i-1}\) and  \(\lamb_{i} \) are distributed according to the Wigner Surmise,  the resulting distribution between \( \eps_{i} \) and  \(\lamb_{i} \),
\begin{equation}
\ps(s,\cin)=\pw(s+\cin)/ \mathcal{N}(\cin)
\qquad\text{with}\quad
\mathcal{N}(\cin)=
\frac{4}{\pi}\cin\, e^{-\frac{4 \cin^2}{\pi }}+\text{erfc}\left(\frac{2 \cin}{\sqrt{\pi }}\right)\,,
\end{equation}
is the sought approximation for \( \pin(s) \) (for \(s\!\geq\!0\)).  Here  \( \mathcal{N}(\cin) \) is fixed by the normalisation condition and   the optimal  value of  \(\cin(\nu)\) has yet to be determined.  Similarly, we can look at the splitting distribution \( \pex(s) \) between \( \lamb_{i} \) and \( \eps_{i+1} \), which we can express based on the same \( \ps(s) \) but with another cutting value \( \cex(\nu) \).

To identify the correct threshold \(c\)'s,  we  demand  \(\ps(0,\cin)\teq \pin(0) \), as well as \(\ps(0,\cex) \teq \pex(0) \).  Using the exact solutions for \(s\teq 0\) from the analytical calculations in Section \ref{sec:analyt} we obtain:
\begin{align}
\pin(0)& =1+\frac{1}{\nu}-\frac{1}{2\pi\nu}\Tr{\Lambda^{-1}}\,,
\\
\pex(0)& = 1-\frac{1}{2\pi\nu}\Tr{\Lambda^{-1}}\,,
\end{align}
see eqs. (\ref{eq:pinFinal}, \ref{eq:pexFinal}) and, for the definition of \( \Lambda \), (\ref{eq:lambdaDef}). Although further analytic treatment of \( \Tr\Lambda^{-1} \) is  possible, it is more convenient to use the  \(R_2(s) \) function  calculated in \cite{aleiner}  which is  depicted in Figure \ref{pic:r2comp}. Recalling that  \(R_2(0) \teq \pin(0) \) holds we obtain
\begin{equation}
\lim_{N\to\infty}\Tr\Lambda^{-1}=2\arctan{\pi\nu}\,.
\end{equation}
This provides us with the necessary relation to determine  \(\cin\), \(\cex\) from  \( \nu\) analytically. For instance, in the limit \(\nu\!\to\!\infty\), \(\norm(\cin)\teq\pw(\cin)\) (likewise for \(\cex\)) gives \(\cin\teq\cex\approx 0.641\).

\begin{figure}[bt]
\centering
\includegraphics[width=0.98\textwidth]{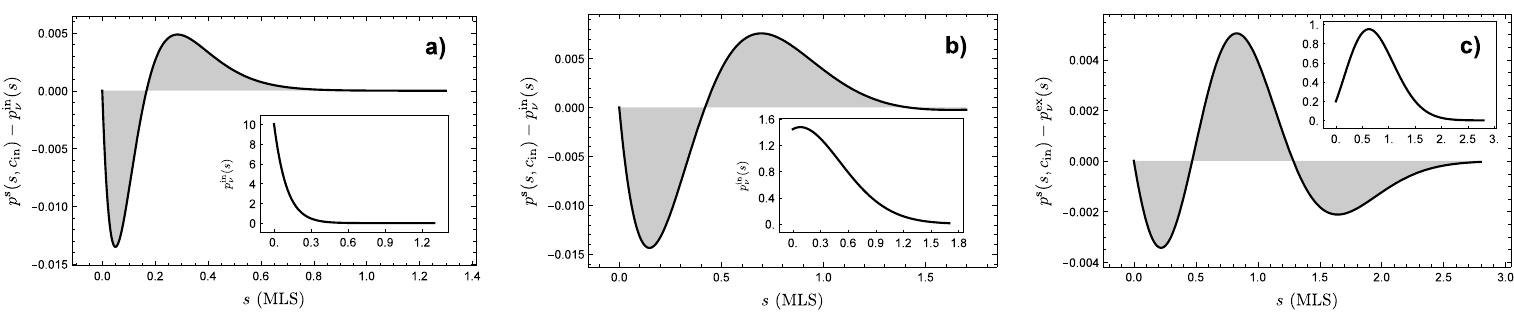}
\caption{Shown are the absolute deviations between the heuristic surmise \( \ps \) and a corresponding analytical curve, which itself is plotted in the inset.
In the cases a) and b) \( \pin(s) \)  is depicted for  \( \nu \teq 0.1 \) and \( \nu\teq 1.3 \). The last panel displays  \( \pex(s) \) for \( \nu\teq 0.3 \). While the error stays relatively constant for \( \pin(s) \), in this case it decreases as  \( \nu\to 0 \), whereat \(\pex(s) \) approaches the GUE nearest neighbour distribution. Due to the discrepancy between the  exact  nearest neighbour distribution  and \( \pw(s) \) the error  will not vanish completely though. MLS is adjusted such that \(\bar{\rho}^{(\epsilon)}\teq 1 \).}
\label{pic:heuristicComp}
\end{figure}

Surprisingly this simple surmise shows a very  good agreement  with the exact result for all ranges of perturbation strength.  The comparison with  the analytics is shown in fig.~\ref{pic:heuristicComp}. 
\section{Conclusion and Outlook}
\label{sec:conclusion}

The main part of this paper is devoted to the analytical calculation of the  nearest neighbour distribution \( \pnu(s) \)
for the spectra of nearly unidirectional quantum graphs. Furthermore, based on the Wigner distribution for GUE we were able to obtain a simple surmise giving a good approximation for \( \pnu(s) \) valid for an arbitrary perturbation strength \(\nu\).  These  results show an excellent agreement with the data  obtained from numerical  calculations for generic (\eg fully connected) graphs. However, for some classes of graphs essential  deviations were found if the perturbation is placed at edges belonging to short loops. It was demonstrated that such deviations can be attributed to a strong scarring effect at these edges.

To investigate this scarring effect further it would be instructive to develop a semi-classical approach based on periodic orbit theory. This would allow the derivation of non-universal corrections to the RMT result based on the specific properties of the graph's edges. The semi-classical approach is also needed to address the question of the splitting distribution in the spectrum of  unidirectional billiard systems such as the Reuleaux polygons. As in the case of graphs, it might be expected that spectral deviations from standard statistics of GUE arise here due to the presence of diffractive orbits. While in graphs such orbits are caused by backscattering  at specially designed vertices, in  the unidirectional billiards the same role could be played by singular classical orbits hitting the billiard's corners.

\begin{figure}[hbtp]
\centering
\includegraphics[width=0.9\textwidth]{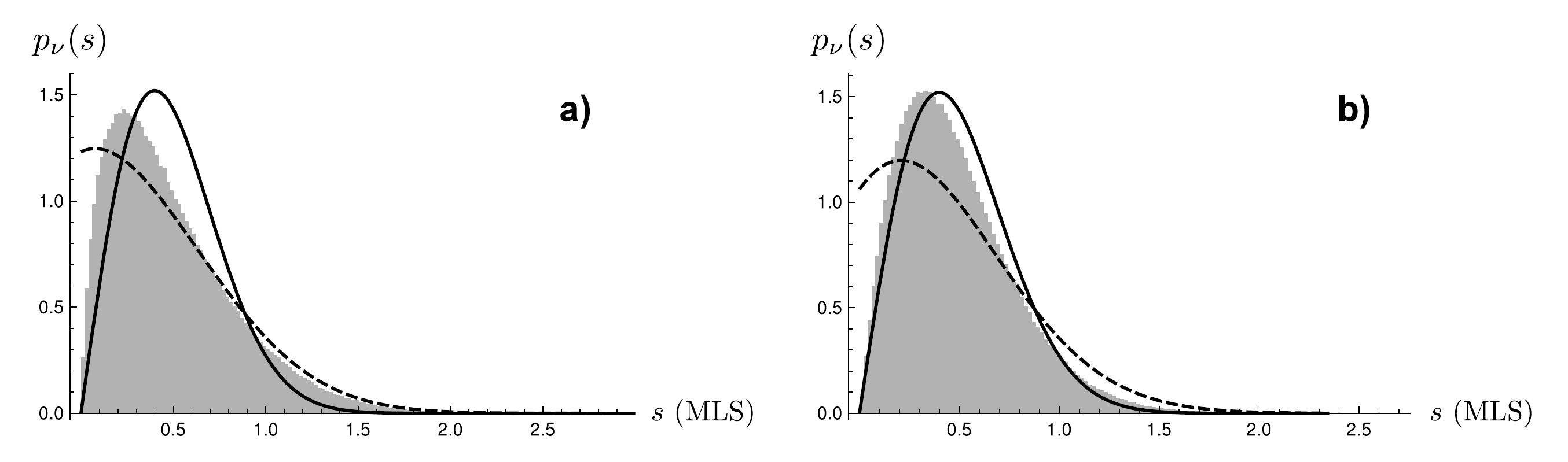}
\caption{The images show the nearest neighbour distribution for a De Bruijn graph with 64 regular vertices and 2 (a) or 4 (b) scatterers placed on different, non-anomalous edges. Even for one additional scatterer (a) we see a strong deviation from our rank 1 result (dashed line) and instead find similarity to a GOE distribution (continuous line) which increases with larger numbers of scatterers. In both cases we find strict level repulsion as \(\pnu(0) \teq 0\).}
\label{pic:HRpert}
\end{figure}

So far, our results were restricted to rank one perturbations. However, for billiards an effective rank of ``perturbation'' should (at least) depend on the number of corners.  As the rank of the perturbation increases it is expected that the originally double degenerate GUE distribution must transform to the non-degenerate GOE. As a small outlook on this transition we numerically determined the nearest neighbour distribution for a graph with two/four backscattering vertices  (effectively a rank two/four perturbation), presented in  fig.~\ref{pic:HRpert}. Note that, besides loosing the interlacing property and the split up into two sub-spectra, we  find that the nearest neighbour distribution vanishes as \(s\to 0\). This immediately implies level repulsion among energy levels of the system, which is absent for rank one perturbations.

\subsection*{Acknowledgements}
This work was supported by the DFG within the Sonderforschungsbereich TR/12 and the research grant Gu 1208/1-1.  We are grateful  to T.~Guhr, V.~Osipov  and  T.~Wirtz for instructive and stimulating discussions.
\appendix
\section*{Appendix}
\section{Calculation of \(F_{kl}\)}
\label{sec:calcF}

For the calculation of the double integrals in equation (\ref{eq:fklInt}) it is convenient to expand the brackets and treat the four arising parts separately.
Let us denote them by \(F_{kl}\teq\Foo-\Fto-\Fot+\Ftt\), where \eg \(\Fto\) stands for the integral over the \(\theta(\eps-\xe)\theta(\ye-\eps)\) part only.
Further on, to compactify the resulting expressions, we introduce \(\kt\teq k-1-(N-1)/2\) and \(\lt\teq l -1-(N-1)/2\). They are given by:
\begin{align}
\Foo=&\frac{2\eu^{\nl\pi}}{\nk\nl}\sinh{\left(\nk \pi\right)}
-\frac{2\pi}{\nl}\delta_{\kt+\lt}\,,\\
\label{eq:fkllamb}
\Fot=&\frac{2}{\nk}\frac{\sin{\left((\kt+\lt)\frac{\yl-\xl}{2}\right)}}{\kt+\lt}\eu^{\iu(\kt+\lt)(\yl+\xl)/2}\\
&-\frac{2\eu^{-\nk\pi}}{\nk\nl} \eu^{\nl\frac{\yl+\xl}{2}}
\sinh{\left(\frac{\nl}{2}(\yl-\xl)\right)}
\,,\nonumber\\
\Fto=& \frac{2\eu^{\nl\pi}}{\nk \nl} \eu^{\nk(\ye+\xe)/2}\sinh{\left(\frac{\nk}{2} (\ye-\xe)\right)} \\
&-\frac{2}{\nl}\frac{\sin{\left((\kt+\lt)\frac{\ye-\xe}{2}\right)}}{\kt+\lt}\eu^{\iu(\kt+\lt)(\ye+\xe)/2}
\,,\nonumber
\end{align}
wherein \(\delta_{\kt+\lt}\) stands for the Kronecker-Delta which is \(1\) if \(\kt+\lt\teq 0\) and zero otherwise while \(\nl\teq\iu\lt-N/(2\nu) \) and \(\nk\teq\iu\kt+N/(2\nu) \). In the case where all four \(\theta\)-functions appear the general solution is slightly more complicated. Taking into account the position of the gap, see figure \ref{pic:gapSpec}, it takes on the form:
\begin{equation}
\begin{split}
\Ftt=& 
\frac{2}{\nk\nl} \eu^{\nl\lmax} \sinh{\left( \nk \frac{\pi}{N}s\right)}
-\frac{2}{\nl}\frac{\sin{\left( (\kt+\lt) \frac{\pi}{N}s\right)}}{\kt+\lt}
\\
+& 2\,\theta{(\lmin-\emin)}
\frac{\eu^{\nl(\lmin+\lmax)/2}}{\nk\nl}
\left( \eu^{\nk\lmin}-\eu^{\nk\emin} \right)
\sinh{\left(\nl \frac{\lmax-\lmin}{2} \right) }\,,
\end{split}
\label{eq:fklepslamb}
\end{equation}
wherein the factor \(\pi/N\) stems from the scaling of the gap boundaries. The Heaviside \(\theta\)-function distinguishes between the case of the internal splitting distribution, where it is one, and the external splitting distribution, where it is zero. Please observe that equation (\ref{eq:fklepslamb}) holds only for the numerator, compare equation (\ref{eq:InDistr}). In the case of the denominator \(\Ftt\teq 0\).

To obtain the expression for the numerator (in the case of internal splitting), see equation (\ref{eq:InDistr}), we set \(\emin\teq -s\pi/N-\dte\), \(\lmin\teq -s\pi/N\), \(\emax\teq +s\pi/N\) and \(\lmax\teq +s\pi/N+\dtl\).\footnote{For the external splitting set: \(\emin\teq -s\pi/N\), \(\lmin\teq -s\pi/N-\dtl\), \(\emax\teq s\pi/N+\dte\) and \(\lmax\teq s\pi/N\)} Furthermore, we are only interested in the case where \(N\gg 1\) which allows for some minor simplifications. After expansion of the equations (\ref{eq:fkllamb}-\ref{eq:fklepslamb}) up to the order \(\dte\dtl\) we obtain:
\begin{equation}
\begin{split}
F_{kl}^\text{(num)}
\approx &\eu^{\iu(\kt+\lt)\pi}-2\pi \left(\iu\kt+\spert\right)\delta_{\kt+\lt}
-2\left(\iu\lt-\spert\right)\frac{\sin{\left((\kt+\lt)\frac{\pi}{N}s\right)}}{\kt+\lt}
\\
&+2\eu^{(\iu\lt-\spert)\sfrac{\pi}{N}\,s} \sinh{\bigg(\left(\iu\kt+\spert\right)\frac{\pi}{N}\,s\bigg)}
+\eu^{\iu(\lt-\kt)\sfrac{\pi}{N}\,s-2\frac{\pi s}{N \nu}}
\\
&\quad\times\bigg(
\left(\iu\kt+\spert\right)\dte- \left(\iu\lt-\spert\right)\dtl+ \left(\iu\lt-\spert\right) \left(\iu\kt+\spert\right)
\dte\dtl \bigg)\,.
\end{split}
\label{eq:fklNumerator}
\end{equation}
In the case of the denominator we choose \(\emin\teq -s\pi/N-\dte\), \(\emax\teq -s\pi/N\) and \(\lmin\teq\lmax\teq 0\).
Applying the same limit and expansions as for the numerator we arrive at:
\begin{equation}
F_{kl}^\text{(den)}\approx
\eu^{\iu(\kt+\lt)\pi} -2\pi\left(\iu\kt+\spert\right)\delta_{\kt+\lt}
+\eu^{-\iu(\kt+\lt)\sfrac{\pi}{N}\,s}
\left(\iu\kt+\spert\right)\dte\,.
\label{eq:fklDenominator}
\end{equation}
Note that the alternating sign \(\eu^{\iu(\kt+\lt)\pi}\) appearing in both \(F_{kl}\) (\ref{eq:fklNumerator},\ref{eq:fklDenominator}) does not affect the result of the determinant and can be neglected. With the notation introduced in section \ref{sec:analyt:split} we may cast both \(F_{kl}\) into the matrix forms given by equations (\ref{eq:fDenom}) and (\ref{eq:fNum}).


\bibliographystyle{ieeetr}
\bibliography{refs}

\begin{thebibliography}{10}

\bibitem{dyson}
F.~J. Dyson, ``{Statistical Theory of the Energy Levels of Complex Systems.
  I},'' {\em Journal of Mathematical Physics}, vol.~3, no.~1, pp.~140--156,
  1962.

\bibitem{berry}
M.~Robnik and M.~V. Berry, ``{False time-reversal violation and energy level
  statistics: the role of anti-unitary symmetry},'' {\em Journal of Physics A:
  Mathematical and General}, vol.~19, no.~5, p.~669, 1986.

\bibitem{AnomalousSpectrum}
F.~Leyvraz, C.~Schmit, and T.~H. Seligman, ``{Anomalous spectral statistics in
  a symmetrical billiard},'' {\em Journal of Physics A: Mathematical and
  General}, vol.~29, no.~22, p.~L575, 1996.

\bibitem{sebastian}
C.~H. Joyner, S.~M{\"u}ller, and M.~Sieber, ``{Semiclassical approach to
  discrete symmetries in quantum chaos},'' {\em Journal of Physics A:
  Mathematical and Theoretical}, vol.~45, no.~20, p.~205102, 2012.

\bibitem{boris1}
B.~Gutkin, ``{Dynamical 'breaking' of time reversal symmetry},'' {\em Journal
  of Physics A: Mathematical and Theoretical}, vol.~40, no.~31, p.~F761, 2007.

\bibitem{boris2}
B.~Gutkin, ``{Note on converse quantum ergodicity},'' {\em Proc. Amer. Math.
  Soc.}, vol.~137, no.~8, pp.~2795--2800, 2009.

\bibitem{prosen}
G.~Veble, T.~Prosen, and M.~Robnik, ``{Expanded boundary integral method and
  chaotic time-reversal doublets in quantum billiards},'' {\em New Journal of
  Physics}, vol.~9, no.~1, p.~15, 2007.

\bibitem{bDietz}
B.~Dietz, T.~Guhr, B.~Gutkin, M.~Miski-Oglu, and A.~Richter, ``{Spectral
  properties and dynamical tunneling in constant-width billiards},'' {\em Phys.
  Rev. E}, vol.~90, p.~022903, Aug 2014.

\bibitem{mSieber}
M.~Sieber, ``{Spectral statistics in chaotic systems with a point
  interaction},'' {\em Journal of Physics A: Mathematical and General},
  vol.~33, no.~36, p.~6263, 2000.

\bibitem{bogomolnyScatterer}
E.~Bogomolny, P.~Leboeuf, and C.~Schmit, ``{Spectral Statistics of Chaotic
  Systems with a Pointlike Scatterer},'' {\em Phys. Rev. Lett.}, vol.~85,
  pp.~2486--2489, Sep 2000.

\bibitem{graphsReview}
S.~Gnutzmann and U.~Smilansky, ``{Quantum graphs: Applications to quantum chaos
  and universal spectral statistics},'' {\em Advances in Physics}, vol.~55,
  no.~5-6, pp.~527--625, 2006.

\bibitem{qugaboo}
G.~Berkolaiko and P.~Kuchment, {\em {Introduction to Quantum Graphs}}, vol.~186
  of {\em {Mathematical Surveys and Monographs}}.
\newblock AMS, 2013.

\bibitem{uzy}
T.~Kottos and U.~Smilansky, ``{Quantum Chaos on Graphs},'' {\em Phys. Rev.
  Lett.}, vol.~79, pp.~4794--4797, Dec 1997.

\bibitem{aleiner}
I.~L. Aleiner and K.~A. Matveev, ``{Shifts of Random Energy Levels by a Local
  Perturbation},'' {\em Phys. Rev. Lett.}, vol.~80, pp.~814--816, Jan 1998.

\bibitem{haake}
F.~Haake, {\em {Quantum Signatures of Chaos}}.
\newblock {Springer Series in Synergetics}, Springer, 3~ed., 2010.

\bibitem{portThomas}
C.~E. Porter and R.~G. Thomas, ``{Fluctuations of Nuclear Reaction Widths},''
  {\em Phys. Rev.}, vol.~104, pp.~483--491, Oct 1956.

\bibitem{simon}
F.~M. Marchetti, I.~E. Smolyarenko, and B.~D. Simons, ``{Universality of
  parametric spectral correlations: Local versus extended perturbing
  potentials},'' {\em Phys. Rev. E}, vol.~68, p.~036217, Sep 2003.

\bibitem{hentschel}
M.~Hentschel, D.~Ullmo, and H.~U. Baranger, ``{Fermi edge singularities in the
  mesoscopic regime: Anderson orthogonality catastrophe},'' {\em Phys. Rev. B},
  vol.~72, p.~035310, Jul 2005.

\bibitem{Baecker}
A.~{B{\"a}cker}, ``{Numerical Aspects of Eigenvalue and Eigenfunction
  Computations for Chaotic Quantum Systems},'' in {\em {The Mathematical
  Aspects of Quantum Maps}} (M.~D. {Esposti} and S.~{Graffi}, eds.), vol.~618
  of {\em {Lecture Notes in Physics, Berlin Springer Verlag}}, pp.~91--144,
  2003.

\bibitem{berkolaiko}
G.~Berkolaiko and B.~Winn, ``{Relationship between scattering matrix and
  spectrum of quantum graphs},'' {\em Trans. Amer. Math. Soc.}, vol.~362,
  no.~12, pp.~6261--6277, 2010.

\bibitem{boris3}
B.~Gutkin and V.~A. Osipov, ``{Clustering of periodic orbits in chaotic
  systems},'' {\em Nonlinearity}, vol.~26, no.~1, p.~177, 2013.

\bibitem{scars}
H.~Schanz and T.~Kottos, ``{Scars on Quantum Networks Ignore the Lyapunov
  Exponent},'' {\em Phys. Rev. Lett.}, vol.~90, p.~234101, Jun 2003.

\bibitem{notDaniel}
S.~Gnutzmann, H.~Schanz, and U.~Smilansky, ``{Topological Resonances in
  Scattering on Networks (Graphs)},'' {\em Phys. Rev. Lett.}, vol.~110,
  p.~094101, Feb 2013.

\end{thebibliography}
\end{document}